\documentclass[journal]{IEEEtran}

\usepackage{amsfonts}
\usepackage[algo2e]{algorithm2e}
\usepackage{xcolor}
\usepackage{graphicx}
\usepackage{amsmath}
\usepackage{newpxtext,newpxmath}
\usepackage{algorithm,algpseudocode,caption}
\ifCLASSINFOpdf
  
\else

\fi

\begin{document}

\title{Disturbance Decoupling and Instantaneous Fault Detection in Boolean Control Networks}

\author{S Sutavani, K Sonam  \textit{Student Member, IEEE}, S Wagh, \textit{Senior Member, IEEE}, N Singh, \textit{Member, IEEE} \newline EED, VJTI, Mumbai, India 
\thanks{S Sutavani is with the Electrical Engineering Department, V. J. T. I., Mumbai, India. e-mail: sutavanisarang@gmail.com}
}


\maketitle

\begin{abstract}
The literature available on disturbance decoupling (DD) of Boolean control network (BCN) is built on a restrictive notion of what constitutes as disturbance decoupling. The results available on necessary and sufficient conditions are of limited applicability because of their stringent requirements. This work tries to expand the notion of DD in BCN to incorporate a larger number of systems deemed unsuitable for DD. The methods available are further restrictive in the sense that system is forced to follow trajectory unaffected by the disturbances rather than decoupling disturbances while the system follows its natural course. Some sufficient conditions are provided under which the problem can be addressed. This work tries to establish the notion of disturbance decoupling via feedback control, analogous to the classical control theory. This approach though, is not limited to DD problems and can be extended to the general control problems of BCNs. Determination of observability, which is sufficient for the fault detection, is proven to be NP-hard for Boolean Control Network. Algorithms based on reconstructability, a necessary condition, of BCN turn out to be of exponential complexity in general. In such cases it makes sense to search for the availability of some special structure in BCN that could be utilized for fault detection with minimal computational efforts. An attempt is made to address this problem by introducing instantaneous fault detection (IFD) and providing necessary and sufficient conditions for the same. Later necessary and sufficient conditions are proposed for solving the problem of instantaneous fault detection along with disturbance decoupling using a single controller.
\end{abstract}


%
\IEEEpeerreviewmaketitle

\section{Introduction}
\IEEEPARstart{B}{oolean} Network ($BN$) and Boolean Control Network ($BCN$) model the systems with binary variable dynamics accurately. Binary logic systems (rather, systems that are completely described with binary logic) can be analysed using $BN$s or $BCN$s. More importantly the system that, at some level of abstraction, behave similar to a logical system can also be modelled and analysed as the $BN$ and/or the $BCN$ e.g. $BCN$ has proven to be a useful tool for modelling in the areas like systems biology, especially the genetic regulatory networks after the pioneering work of S. Kauffman in \cite{kauffman1969metabolic}, in which randomly constructed molecular automata was examined by modeling the gene as a binary (on-off) device to study the behavior of large, randomly constructed nets of binary genes. Moreover, $BN$ and $BCN$ may be useful heuristic tools to sense possible behaviours of the systems too complicated to analyse or simulate \cite{sutavani2018interpretation}.

Advances on various control theoretical frontiers of the $BN$ and $BCN$, over the last decade or so, can be primarily attributed to the discrete-time linear dynamical representation of the same developed by D. Cheng and co-authors. \cite{cheng2010linear} introduced linear expression of logic form with Boolean network ($BN$) represented as discrete-time linear system for the investigation of the topological structure. In \cite{cheng2011identification} necessary and sufficient conditions were presented for the identification of state equation from input-state data along with numerical algorithm for approximate identification for large size networks. In \cite{cheng2010state} authors provided  a comprehensive framework for the state–space analysis of Boolean networks by introducing the state space (topological space) and its subspaces of a $BN$ particularly, the regular subspace, the Y-friendly subspace and the invariant subspace. \cite{cheng2011stability} commented on stability and stabilization of BN with necessary and sufficient conditions for constant, open-loop as well as closed-loop control using linear algebraic form and logic coordinate transformation. In \cite{cheng2009controllability} controllability was studied with the help of reachable sets and a necessary and sufficient condition was proposed for observability. \cite{laschov2012controllability} introduced two definitions for controllability of a $BCN$ and provided necessary and sufficient conditions for both the forms in terms of known results from the Perron-Frobenius theory. \cite{li2014controllability} discussed reachability and controllability with system trajectory restricted to avoid undesirable states set providing necessary and sufficient condition for the same. Analytical investigations of reachability and controllability of BCNs with pinning controllers was presented in \cite{lu2015pinning} along with number of possible controllers. \cite{li2018equivalence} investigated the equivalence between state feedback controller and free sequence controller in $BCN$ commenting on implication relation for stabilizability and controllability with respect to the two control structures. \cite{fornasini2012observability} provided necessary and sufficient conditions for observability and reconstructability to hold. \cite{fornasini2012stateobservers} introduced and provided complete characterization of observability and reconstructability properties of BNs and BCNs based both on corresponding matrix representation and network digraph. Further, the problem of state observer design for reconstructible $BN$s and $BCN$s was addressed with proposal of two solutions. \cite{li2014observability} presented more general necessary and sufficient conditions on observability of a $BCN$ without presuming its controllability. \cite{laschov2013observability} showed that the determining the observability of a $BN$ is NP-hard. \cite{zhang2014observability} suggested a unified framework to judge the observability of the $BCN$ pertaining to various definitions of the of the observability. In \cite{cheng2016note} authors investigated the observability of $BCN$ by classifying state pairs into three classes and providing necessary and sufficient conditions. In \cite{zhang2016observer} authors suggested recursive methods to check reconstructability and for reconstructible $BCN$ proposed an approach for design of an online implementable Luenberger-like observer. \cite{zhu2018observability} proposed the concept of a reachable set that results in a given set of initial states and derived necessary and sufficient conditions for the observability of $BCN$s. \cite{hochma2013symbolic} suggested all-together different approach of symbolic dynamics for the study of $BCN$ to tackle the problem of exponentially increasing matrix dimension faced by semi-tensor product (STP) based linear representation. \cite{li2015controllability} utilized similar approach to derive necessary and sufficient conditions for reachability, controllability and observability and devised algorithms to check for the same.

Arguably one of the most important metric indicating the quality of any control system is its ability to nullify or to reject the disturbances i.e. disturbance decoupling. Disturbance in general could represent anything from the omitted higher order terms in the model, the uncertainties in the model to the uncertainties in the operating environment. Naturally the better the disturbance decoupling the closer is the behaviour of the system to the desired one. The performance of the $BCN$ in presence of the disturbance has attracted significant interest. Disturbances in linear time-invariant (LTI) systems are characterised to be of low frequency and generally control systems deal with disturbances by means of higher gain in the feedback loop for lower frequency signals usually resulting in, the output appearing as a result of disturbance, $y_d$ tending to $0$ as time $t$ tends to $\infty$ with reference signal $r=0$. \cite{cheng2010disturbance} provided systematic way to construct output-friendly subspace. Under the output-friendly coordinate frame the solvability of disturbance decoupling problem (DDP) was converted to solving a set of algebraic equations, by putting the dynamics of output-related state variables into a variable separated form. A necessary and sufficient condition was obtained for the solvability of DDP and a control design technique was presented. In \cite{yang2013controller} authors derived a necessary and sufficient condition for the solvability of DDP by analyzing the redundant variables under the framework of output-friendly subspace and a proposed a  method to construct all the valid feedback control matrices.  In \cite{li2017event} the DDP of $BCN$s was investigated under event-triggered control and provided control design algorithms. \cite{liu2017pinning} proposed rank-conditions based pinning control for the solution of DDP under a special dynamic structure suggesting pinning output feedback based controllers and proposed pin-node selection scheme. 

\cite{fornasini2015fault} addresses the problem of fault detection in $BCN$ investigating completeness and T-completeness (based on possible behavior of future input-output trajectory) and introducing meaningful fault to provide necessary and sufficient conditions for detection of fault. \cite{ettore2015fault} investigated the on-line and off-line fault detection problems for $BCN$s, by assuming only two possible configurations, a non-faulty and a faulty one and considering fault to be non self-correcting and presented algorithm for both cases. In \cite{li2012boolean} the Boolean derivative calculation was introduced and applied to fault detection of combinational circuits using STP. Complexity of algorithms based on reconstructability increases exponentially with increasing number of Boolean variables. Looking for some special structure in $BCN$ that could reduce the computational efforts can be of grate importance for fault detection of certain critical systems. In view of this problem, a structure of matrix form defining the $BCN$ is characterized for instantaneous fault detection. 

In literature no attempt has been made to tackle the problems of disturbance decoupling and fault detection simultaneously. In sequel necessary and sufficient conditions are proposed for solving the problem of instantaneous fault detection and disturbance decoupling simultaneously using a single controller. The presence of the special structure not only allows for instantaneous fault detection but also fits nicely with controller structure useful for disturbance decoupling making it possible to design a single controller that can do the job of disturbance decoupling (both, standard as per the literature and the theory presented in this work) as well as fault detection (instantaneous).

\section{Preliminaries}
Boolean dynamical system restricts the variables involved, to a domain defined by the set $\mathcal{D}:= \{ 0, 1\}$. A Boolean function of $n$ variables, $f_B:\mathcal{D}^n\rightarrow \mathcal{D}$, can be considered to represent a map from $\mathcal{D}^n$ (Cartesian product of $n$ $\mathcal{D}$ sets) to $\mathcal{D}$. Formulation of Boolean systems as linear systems (discrete time) using semi-tensor product (STP) opened up the possibility of utilizing well developed techniques from the linear systems theory as tool for the analysis of Boolean systems. The notations and results utilized in the subsequent sections are presented in the following.

\begin{itemize}
    
\item $\mathcal{M}_{m\times n}$ is the set of $m \times n$ real matrices. When $m = n$, it is briefly denoted as $\mathcal{M}_n$

\item $Col(A)(Row(A)) $ is the set of columns (rows) of A

\item $Col_{i}(A)(Row_{i}(A))$ is the ${i}^{th}$ column (row) of A

\item $\delta_{n}^{i}$ is the ${i}^{th}$ column of the identity matrix $I_{n}$

\item $\Delta_{n}:= Col(I_{n})$

\item $L \in \mathcal{M}_{m\times n}$ is called a logical matrix, if $Col(L)\subset \Delta_{m}$. The set of all $ m \times n $ logical matrices is denoted by $\mathcal{L}_{m \times n}$

\item A logical matrix $ L = \left[ \delta_{m}^{i_{1}} \delta_{m}^{i_{2}} ... \delta_{m}^{i_{n}} \right]$is briefly denoted as $L = \delta_{m}\left[i_{1}, i_{2}, ..., i_{n} \right]$, where $i_j \in \{1, \dots, m\}$

\end{itemize}
\textbf{Definition 1} (Kronecker product): The Kronecker product of the matrix $A\in \mathcal{M}_{p\times q}$, with the matrix $B\in \mathcal{M}_{r\times s}$ is defined as:
\begin{eqnarray}
    A \otimes B = \begin{bmatrix}
                    a_{11}B & \dots & a_{1q}B\\ 
                    \vdots &  & \vdots\\ 
                    a_{p1}B & \dots & a_{pq}B
                  \end{bmatrix}
\end{eqnarray}
\textbf{Definition 2} (Semi-tensor product): Let $A \in \mathcal{M}_{m \times n}$ and $B \in \mathcal{M}_{p \times q}$. Denote by $t := lcm(n,p)$  the least common multiple of n and p, then the semi-tensor product of A and B can be defined as
\begin{eqnarray}
    A\ltimes B := (A \otimes I_{t/n})(B \otimes I_{t/p}) \in {M}_{(mt/n)\times(qt/p)}
\end{eqnarray}
1 and 0 are respectively denoted in vector form as:
\begin{eqnarray}
    1 := \left[ \begin{array}{c} 1 \\ 0 
\end{array} \right], 0:= \left[ \begin{array}{c}
 0 \\ 1
\end{array} \right]
\end{eqnarray}
\textbf{Definition 3} (structure matrix) A $2 \times 2^{r}$ matrix $M_{\sigma} \in \mathcal{L}_{m \times n}$ is said to be the structure matrix of the logical operator $\sigma: \mathcal{D}^r \rightarrow \mathcal{D}$ if 
\begin{eqnarray}
\sigma(p_{1},...,p_{r})	= M_{\sigma} \ltimes p_{1} \ltimes ... \ltimes p_{r} := M_{\sigma} \ltimes_{i=1}^{r} p_{i}
\end{eqnarray}

\begin{table}[]
\centering
\caption{Structure Matrices for Logical Operators}
\label{structure_matrix_table}
\begin{tabular}{|l|l|l|}
\hline
\textbf{Logical Operator} & \textbf{Notation} & \textbf{Structure Matrix}   \\ \hline
Negation  & $\neg$ & $M_{n}$ = $\delta_{2}${[}2 1{]}     \\ \hline
Conjunction & $\vee$ & $M_{c}$ = $\delta_{2}${[}1 2 2 2{]} \\ \hline
Disjunction & $\wedge$ & $M_{d}$ = $\delta_{2}${[}1 1 1 2{]} \\ \hline

\end{tabular}
\end{table}
The structure matrices for some basic Boolean functions (TABLE \ref{structure_matrix_table}) and their representation is as follows:
\begin{eqnarray}
\begin{aligned}
\neg p &= M_{n}p  \\
p \wedge q &= M_{c}p q \\
p \vee q &= M_{d}p q \\
\end{aligned}
\end{eqnarray}
\textbf{Definition 4} (Dummy operator) Dummy operator $E_{du} \in \mathcal{L}_{m \times n}$ is defined as $E_{du}(p,q) := q $, $ \forall p,q \in \mathcal{D} $. Its structure matrix and expression is as follows:
    \begin{eqnarray}
        E_{du} = \left[\begin{array}{cccc} 1 &0 &1 &0 \\ 0 &1 &0 &1
        \end{array}\right]
    \end{eqnarray} 
    \begin{eqnarray}
        E_{du}pq = q
    \end{eqnarray}

Consider a Boolean dynamical system referenced by (\ref{eq_BN_sys}) with $X_i \in \mathcal{D}$ and $f_i: \mathcal{D}^n \rightarrow \mathcal{D}$, 
 \begin{eqnarray} \label{eq_BN_sys}
 \begin{cases} X_{1}(t+1) = f_{1}(X_{1}(t), \dots, X_{n}(t))\\	
 \vdots \\
 X_{n}(t+1) = f_{n}(X_{1}(t), \dots, X_{n}(t))
\end{cases}
\end{eqnarray}  
%
The variables $X_1, \dots, X_n$ are called the state variables of the system. The vector form of the state variables, denoted by $\bar{X}_i$ for $i\in \{1. \dots, n\}$, assumes values from the set $\left \{\begin{bmatrix} 1\\ 0 \end{bmatrix}, \begin{bmatrix} 0\\ 1 \end{bmatrix} \right \}$ i.e. $\bar{X}_i \in \Delta_2$. The uniquely defined quantity $x = \bar{X}_1 \ltimes \bar{X}_2 \ltimes \dots \ltimes \bar{X}_n = \ltimes_{i=1}^{n} \bar{X}_i$ is called the state of the system.  This system of equations defines a $BN$. System described by (\ref{eq_BN_sys}) also has representation in the form of a digraph of a finite automaton.\\
\textbf{Corollary 5} \cite{cheng2010linear} System (\ref{eq_BN_sys}) can be expressed in linear form as
    \begin{eqnarray} \label{eq_BN_linear}
        x(t+1) = Lx(t)
    \end{eqnarray}
where $ x = \ltimes_{i=1}^{n} \bar{X}_{i}$, a unique state transition matrix of the system $ L \in \mathcal{L}_{2^{n} \times 2^{n}}$, such that $L = M_{f_1}\ltimes M_{f_2}\ltimes ...\ltimes M_{f_n}$ and $M_{f_i}\in \mathcal{L}_{2 \times 2^{n}}$ $\forall i$, is the structure matrix of $f_i: \mathcal{D}^n \rightarrow \mathcal{D}$. Equation (\ref{eq_BN_linear}) is called the algebraic form of system (\ref{eq_BN_sys}).

In a Boolean system, if a variable (dynamics) is not influenced by any variable of the (including itself) then it is called as the input variable (there could be multiple input variables). If a variable does not affect any variable (dynamics) of the system, then it is termed as the output variable. A Boolean network with both input and output variables is called as a Boolean Control Network. The system of equations (\ref{eq_BCN_sys_state}), (\ref{eq_BCN_sys_output}) represents a $BCN$.
\begin{eqnarray}\label{eq_BCN_sys_state}
 \begin{cases}
    X_{1}(t+1) = f_{1}(X_{1}(t), \dots, X_{n}(t), U_{1},..., U_{m})\\	
    \vdots \\
    X_{n}(t+1) = f_{n}(X_{1}(t), \dots, X_{n}(t), U_{1},\dots, U_{m})
 \end{cases}
\end{eqnarray} 
\begin{eqnarray}\label{eq_BCN_sys_output}
 \begin{cases}
    Y_{1}(t) = h_{1}(X_{1}(t),\dots, X_{n}(t))\\
    \vdots \\
    Y_{p}(t) = h_{p}(X_{1}(t),\dots, X_{n}(t))
 \end{cases}
\end{eqnarray} 
Where, $X_i, U_j, Y_k \in \mathcal{D}$ and $f_i: \mathcal{D}^{n+m} \rightarrow \mathcal{D}$, $g_k: \mathcal{D}^{n} \rightarrow \mathcal{D}$. $X_1, \dots, X_n$ are the state variables, $U_1, \dots, U_m$ are the input variables and $Y_1, \dots, Y_p$ are the output variables of the system. The corresponding vector forms, denoted by $\bar{X}_i$ for $i\in \{1. \dots, n\}$, $\bar{U}_j$ for $j\in \{1. \dots, m\}$ and $\bar{Y}_k$ for $k\in \{1. \dots, p\}$ respectively, assume values from the set $\left \{\begin{bmatrix} 1\\ 0 \end{bmatrix}, \begin{bmatrix} 0\\ 1 \end{bmatrix} \right \}$ i.e. $\bar{X}_i,\bar{U}_j,\bar{Y}_k \in \Delta_2$. $x = \bar{X}_1 \ltimes \bar{X}_2 \ltimes \dots \ltimes \bar{X}_n = \ltimes_{i=1}^{n} \bar{X}_i$, $u = \bar{U}_1 \ltimes \bar{U}_2 \ltimes \dots \ltimes \bar{U}_m = \ltimes_{j=1}^{m} \bar{U}_j$ and $y = \bar{Y}_1 \ltimes \bar{Y}_2 \ltimes \dots \ltimes \bar{Y}_p = \ltimes_{k=1}^{p} \bar{Y}_k$ are the state, the input and the output of the system respectively.\\
\textbf{Corollary 6} System defined by (\ref{eq_BCN_sys_state}) and (\ref{eq_BCN_sys_output}) can be expressed in linear form as
\begin{eqnarray} \label{eq_BCN_linear_state}
    x(t+1) &=& Lu(t)x(t)\\ \label{eq_BCN_linear_output}
    y(t) &=& Hx(t) 
\end{eqnarray}
where $x = \ltimes_{i=1}^{n} \bar{X}_{i}$, $u = \ltimes_{j=1}^{m} \bar{U}_{j}$, $ y = \ltimes_{k=1}^{p} \bar{Y}_{k}$, $ L \in \mathcal{L}_{2^{n} \times 2^{n+m}}$ such that $L = M_{f_1}\ltimes M_{f_2}\ltimes ...\ltimes M_{f_n}$ is the unique state transition matrix, $M_{f_i}\in \mathcal{L}_{2 \times 2^{n+m}}$ $\forall i$ is the structure matrix of $f_i: \mathcal{D}^{n+m} \rightarrow \mathcal{D}$, $H \in \mathcal{L}_{2^{p} \times 2^{n}}$ such that $H = M_{h_1}\ltimes M_{h_2}\ltimes ...\ltimes M_{h_p}$ and $M_{h_i}\in \mathcal{L}_{2 \times 2^{n}}$ $\forall i$ is the structure matrix of $h_i: \mathcal{D}^n \rightarrow \mathcal{D}$. Equation (\ref{eq_BCN_linear_state}) is the algebraic form of the system (\ref{eq_BCN_sys_state}) and equation (\ref{eq_BCN_linear_output}) is the algebraic form of the system (\ref{eq_BCN_sys_output}).

A $BCN$ with $n$ state variables and $p$ output variables, has $2^n$ distinct states and $2^p$ distinct outputs respectively. $O_{si}:= \{x_j \in \Delta_{2^n}\ |\ Hx_j = y_i\}\ \forall i \in \{1, \dots, 2^p\}$ i.e. $O_{si}$ is the set of all the states that produce (result in) the output $y_i$.


\section{Behavioural equivalence in BN \& BCN}


Whether to use a $BN$ based or a $BCN$ based model, depends upon the nature of the system as well as the problem related to the same being addressed. Uncertainties with problem definition leaves possibilities for both $BN$ and $BCN$ based models open. In such cases, expecting both the type of models to follow behaviour that is equivalent in some sense is not unreasonable. This idea of behavioural equivalence expressed here can be broadly classified in three categories, namely structural equivalence based on (i) attractor behaviour  (ii) output behaviour (iii) state-transition behaviour. Systems may be termed equivalent in one or more behaviours if the respective behaviours (attractor, output or state-transition) are identical. In relation to continuous systems the $BN$ is analogous to an autonomous system and $BCN$ to a non-autonomous one. The structural equivalence can be looked upon as the existence of the possibility for the external signals to the non-autonomous systems under the influence of which, it behaves similar to an autonomous system. It is clear that, since the $BN$ behaviour can not be influenced externally, for structural equivalence of any kind the $BCN$ should induce the same behaviour under some input.

\subsection {Behavioural Equivalence}
\subsubsection {Behavioural equivalence based on the attractor behaviour}
A BCN is said to be behaviourally equivalent to a BN under the attractor behaviour if under some input scheme the BCN has the same attractors as that of BN and starting from any initial condition the same attractor in BN and BCN is reachable i.e. if the attractors of the BN are represented as 
\begin{equation}
    A_i = \{x_i^1, x_i^2,....,x_i^{k_i}\} \quad \quad for\quad 1\leq i\leq N
\end{equation}
where, $N$ - number of attractors of BN, $K_i$ - length of the $i^{th}$ attractor, for a BCN with state transition matrix defined by $L$, $Lx_i^l$ = $x_i^{l+1}$ \& $Lx_i^{k_i}$ = $x_i^1$, the attractors of the BCN are represented as $\bar{A}_i^{\bar{u}}$ = ${\left \{ \bar{x}_j^1, \bar{x}_j^2,...,\bar{x}_j^{\bar{k}_j} \right \}}$ for $j\in  \{1,\dots,M_{\bar{u}}\}$ where, $M_{\bar{u}}$ - number of attractors of the BCN under input scheme $\bar{u}$. Then the two are behaviourally equivalent if \\
$N = M_{\bar{u}}$, $A_m = \bar{A}_m^{\bar{u}}$ for $m\in  \{1,\dots,N/M_{\bar{u}}\}$ and $\mathcal{A}_o(x_i) = \mathcal{A}_o(\bar{x}_i)\ \forall x_i\equiv \bar{x}_i$, where $\mathcal{A}_o(\cdot):-$ indicates attractor of the argument $x_i/\bar{x}_i$ i.e. a map from initial state $x_i/\bar{x}_i$ to the corresponding attractor in BN/BCN under fixed input strategy and $x_i/\bar{x}_i-$ state of BN/BCN respectively.

\subsubsection {Equivalence based on output sequence}
Sometimes it may be the case that what is of interest is only what can be observed, i.e. only the output of the system. A BCN is behaviourally equivalent in output to a BN, if under some input scheme the BCN has the output sequence exactly the same as that of the BN, starting from every initial condition. If starting from any initial state $x_0^i$ the BN produces the output sequence $Y_i = y_i^1 \rightarrow  y_i^2 \rightarrow  ... \rightarrow  y_i^f \rightarrow y_{ia}^1 \rightarrow y_{ia}^2 \rightarrow ... \rightarrow y_{ia}^{k_a} \rightarrow  y_{ia}^1 \rightarrow y_{ia}^2 \rightarrow ... \rightarrow y_{ia}^{k_a}...$ where, $y_i^j$ for $1\leq j\leq f$ is the free output sequence before the system trajectory enters any of the attractors such that $f$ is the distance of $x_o^i$ from the attractor. $y_{ia}^l$ for $1\leq l\leq K_a$ the periodic output sequence after trajectory enters the attractor $a$ with $K_a$, the length of the attractor. If starting from initial state $\bar{x}_o^i$ under the influence of control scheme $\bar{u}$ the BCN produces the output sequence $\bar{Y}_{\bar{u}i} := y_i^{\bar{u}1}, y_i^{\bar{u}2}, y_i^{\bar{u}3},..., y_i^{\bar{u}f_{\bar{u}}}, y_{i\bar{a}}^{\bar{u}1},..., y_{i\bar{a}}^{\bar{u}k_{\bar{a}}}, y_{i\bar{a}}^{\bar{u}1},..., y_{i\bar{a}}^{\bar{u}k_{\bar{a}}},... $ where, $y_i^{\bar{u}j}$, for $1\leq j \leq f_{\bar{u}}$, is free output sequence before the system trajectory enters any of the attractors, $f_{\bar{u}}$ is the distance (of $\bar{x}_o^i$) from the attractor. $y_{i\bar{a}}^{\bar{u}l}$ for $1\leq l\leq \bar{k}_{\bar{a}}$, the periodic output sequence after trajectory enters attractor $a$, where $\bar{k}_{\bar{a}}$ - length of the attractor. Then the BCN is said to be behaviourally equivalent in output sequence to a BN if for every initial state $x_i$ the two output sequences are the same i.e. $Y_i = Y_{\bar{
u}i}$. Note that the first output of $Y_{\bar{u}i}$ i.e. $y_i^{\bar{u}1}$ is independent of $\bar{u}$, $\bar{u}$ is only used for the consistency of the representation.

This equivalence can be modified slightly by comparing the attractor output cycles, i.e. $Y_i^{\infty} = \{ y_{ia}^1\rightarrow y_{ia}^2\rightarrow ...\rightarrow y_{ia}^{k_a}\rightarrow y_{ia}^1 \}$ and $\bar{Y}_{\bar{u}i}^{\infty} = \{ y_{i\bar{a}}^{\bar{u}_1}\rightarrow y_{i\bar{a}}^{\bar{u}_2}\rightarrow ...\rightarrow y_{i\bar{a}}^{\bar{u}_{K_{\bar{a}}}}\rightarrow y_{i\bar{a}}^{\bar{u}_1} \}$ represent the same cycle. Equivalence in the output cycle is weaker than equivalence in attractors, as in attractor $\bar{A}_j^{\bar{u}} = \{\bar{x}_j^1, \bar{x}_j^2, ..., \bar{x}_j^{\bar{k}j}\}$ replacing any $\bar{x}_j^i$ for $1\leq i\leq \bar{k}_i$, by any other states from its output set $OS_{\bar{x}_j^i}$ will destroy the equivalence in attractors, but equivalence in output cycle will remain unchanged. 

\subsubsection {Equivalence based on the state transition sequence}
This is the strongest equivalence that can be achieved between a BCN and a BN. A BCN is equivalent to a BN in the state transition sequence, if under some input scheme $\bar{u}$ every state transitions to a state, same as that of in the BN. In other words, the BCN under input scheme $\bar{u}$ transforms to the BN; therefore generates the same network graph and the same state transition matrix $L$ as that of the BN. If $\tilde{L}_{\bar{u}}$ represents the structure of the state transition matrix of the BCN under the input scheme, then the BCN is behaviourally equivalent to the BN in state transition sequence if $\forall x_i \in \Delta_{2^n}$
\begin{equation*}
    Lx_i = \tilde{L}_{\bar{u}}x_i
\end{equation*}
This equivalence is the easiest to verify, as it requires the exact reconstruction of the $L$ matrix under some input scheme. (State feedback is the best scheme possible, in the sense that the decision for control is taken based on the current state, which is sufficient to achieve desired response if possible).
\subsection{Mathematical Analysis}
Consider a BN and a BCN defined by (8) and (9) respectively.
\subsubsection{Behavioural Equivalence Based on State Transition Sequence}
Let $x^+$ denote the next state in time for both BN and BCN, therefore 
\begin{eqnarray*}
    x^+ &=& x^+\\
    LE_{du}ux &=& \tilde{L}ux
\end{eqnarray*}
where, $E_{du}$ is a dummy for the first variable $u$. If $LE_{du} = \tilde{L}$, then BN is equivalent to BCN in state transition sequence for every possible $u$ i.e. in $\tilde{L}$, $u$ is a dummy variable that does not affect the dynamics. If state feedback is utilized with feedback matrix $M_x$, i.e. $u = M_x x$ then,
\begin{eqnarray*}
    Lx &=& \tilde{L}M_x \psi_nx
\end{eqnarray*}
where, $\psi_n$ is the power reducing matrix. $\tilde{L}\in \mathcal{L}_{m+n}$ is the state transition matrix. $\tilde{L}_{M_x} := \tilde{L}M_x \psi_n \in \mathcal{L}_{n\times n}$ is obtained as\cite{yang2013controller} $\tilde{L}_{M_x}:=\left [ Col_1(Blk_{Col_1 M_x}\tilde{L}) \,\,
... \,\,Col_{2^n}(Blk_{Col_{2^n} M_x}\tilde{L}) \right ]$. If $L=\tilde{L}_{M_x}$ then the BN and the BCN are equivalent for $M_x$ as the state feedback matrix.

For output feedback defined by $M_y$, 
$$Lx=\tilde{L}M_yH\psi_nx$$
If $L=\tilde{L}_{M_y}$ then the BN and the BCN are equivalent for $M_x$ as the output feedback matrix with $\tilde{L}_{M_y}=\tilde{L}M_yH\psi_n$.

\subsubsection{Behavioural Equivalence Based On Output Sequence}
Let $y^+$ denote the next output in time for both BN and BCN, then
\begin{eqnarray*}
    y^+ &=& y^+\\
    HLx &=& H\tilde{L}ux\\
    HLE_{du}ux &=& H\tilde{L}ux
\end{eqnarray*}
where, $H$ is the output matrix of the system. Therefore, if $HLE_{du} = H\tilde{L}$, then $BN$ is behaviourally equivalent to $BCN$ in output sequence for every possible $u$.
\subsubsection* {i) State Feedback}
Consider the case of state feedback represented by $M_x$ then,
\begin{eqnarray*}
   HLx &=& H\tilde{L}M_xS \psi_nx
\end{eqnarray*}
therefore $HL = H\tilde{L}M_x \psi_n$ or $HL = H\tilde{L}_{M_x}$ where $H\tilde{L}_{M_x} = \tilde{L}M_x \psi_n$. If $Col_i(L)\in O_{sj}$ and $Col_i(\tilde{L}_{M_x})\in O_{sj}$ then $BN$ is equivalent to $BCN$ in output sequence for state feedback defined by $M_x$.
\subsubsection*{ii) Output Feedback}
Consider the case of output feedback represented by $M_y$ then,
\begin{eqnarray*}
    HLx &=& H\tilde{L}M_yH \psi_nx
\end{eqnarray*}
therefore if $HL = H\tilde{L}y$, where $\tilde{L}_y := \tilde{L}M_yH\psi_n$ + strcuture of $\tilde{L}_y$ and $Col_i(L)\in O_{sj}$ \& $Col_i(\tilde{L}y)$, then the $BN$ is equivalent to $BCN$ in output sequence for output feedback defined by $M_y$.

\subsubsection{Behavioural Equivalence Based on Attractor Behaviour}
Let $\mathcal{S_A}:=$ \{Set of all the states that belong to some attractor\} i.e. $\mathcal{S_A}:=\{ x_i\ |\ \exists T\geq 0\ \text{such that}\ L^Tx_i=x_i\}$. Let $x_i\in \mathcal{S_A}$ and $x_i^+$ denote its next state in time,
\begin{eqnarray*}
    Lx_i &=& \tilde{L}ux_i\\
    LE_{du}ux_i &=& \tilde{L}ux_i
\end{eqnarray*}
Let, $\mathcal{S}_{\mathcal{A}u}:= \{ u\ltimes x_i\mid x_i\in \mathrm{S_A}\ \& u\in \Delta_{2^m}\}$ a set of Boolean vectors formed by taking STP of control inputs from admissible input set and states from $\mathcal{S_A}$. It can equivalently be defined as $\mathcal{S}_{\mathcal{A}u}:= \{j\in \Delta_{2^{m+n}}\mid j=(2^n)k+1,\ 1\leq k\leq 2^m-1\ \&\  \delta_{2^n}^i\in \mathcal{S_A}\}$. If $Col_{\mathcal{S}_{\mathcal{A}u}}(LE_{du}) = Col_{\mathcal{S}_{\mathcal{A}u}}(\tilde{L})$ and $\mathcal{A}_o(x_i) = \mathcal{A}_o(\bar{x}_i)\ \forall x_i\equiv \bar{x}_i$, then $BN$ is equivalent to $BCN$ for all inputs where, $Col_{\mathcal{S}_{\mathcal{A}u}}(.)$ indicates the set of columns of a logic matrix that indexed by $\mathcal{S}_{\mathcal{A}u}$.

\subsubsection*{i) State Feedback}
Let $x_i\in \mathcal{S_A}$ and $x_i^+$ denote its next state in time,
\begin{eqnarray*}
   Lx_i &=& \tilde{L}M_x\psi_nx_i = \tilde{L}_{x_f}x_i
\end{eqnarray*}
If $Col_{\mathcal{S_A}}(L) = Col_{\mathcal{S_A}}(\tilde{L}x)$ and $\mathcal{A}_o(x_i) = \mathcal{A}_o(\bar{x}_i)\ \forall x_i\equiv \bar{x}_i$ then $BN$ is equivalent to $BCN$ for state feedback defined by $M_x$.

\subsubsection*{ii) Output Feedback}
Let $x_i\in \mathcal{S_A}$ and $x_i^+$ denote its next state in time,
\begin{eqnarray*}
    Lx_i &=& \tilde{L}ux_i\\
    Lx_i &=& \tilde{L}M_yH\psi_nx_i = \tilde{L}_{y_f}x_i
\end{eqnarray*}
If $Col_{\mathcal{S_A}}(L) = Col_{\mathcal{S_A}}(\tilde{L}y)$ then $BN$ is equivalent to $BCN$ for output feedback defined by $M_y$.

\subsubsection{Output Steady State Equivalence}
Let $x_i^+$ and $y_i^+$ denote the next state and output in time of state $x_i$ respectively,
\begin{eqnarray*}
    y_i^+ &=& y_i^+\\
    HLx_i &=& H\tilde{L}ux_i\\
    HLE_{du}ux_i &=& H\tilde{L}ux_i
\end{eqnarray*}
If $Col_j(LE_{du})\in OS_l$, $Col_j(\tilde{L})\in OS_l$ $\forall j\in \mathcal{S}_{\mathcal{A}u}$ for some $1\leq k\leq 2^p$ and $\mathcal{A}_o(x_i) = \mathcal{A}_o(\bar{x}_i)\ \forall x_i\equiv \bar{x}_i$, then $BN$ is equivalent to $BCN$ for all inputs.

\subsubsection*{i) State Feedback}
Let $x_i^+$ and $y_i^+$ denote the next state and output in time of state $x_i$ respectively,
\begin{eqnarray*}
    HLx_i &=& H\tilde{L}M_x\psi_nx_i = H\tilde{L}_xx_i
\end{eqnarray*}
If $Col_j(L)\in OS_k$ and $Col_j(\tilde{L}_x)\in OS_k$ $\forall j\in \mathcal{S}_{\mathcal{A}u}$ and for some $1\leq k\leq 2^p$, then $BN$ is equivalent to $BCN$ for state feedback defined by $M_x$.

\subsubsection*{ii) Output Feedback}
Let $x_i^+$ and $y_i^+$ denote the next state and output in time of state $x_i$ respectively,
\begin{eqnarray*}
    HLx_i &=& H\tilde{L}M_yH\psi_nx_i = H\tilde{L}_yx_i
\end{eqnarray*}
If $Col_j(L)\in OS_k$ and $Col_j(\tilde{L}_y)\in OS_k$ $\forall j\in \mathcal{S}_{\mathcal{A}u}$ and for some $1\leq k\leq 2^p$, then $BN$ is equivalent to $BCN$ for output feedback defined by $M_y$.

This analysis can be further extended for the BN - BN, BN - BCN equivalence in presence of disturbance ($\xi$) under following assumptions:
\begin{itemize}
    \item Disturbance affects only BCN
    \item Disturbance affects both BN and BCN
\end{itemize}
The second case is trivial, as $x\ltimes \xi$ can be considered as the state of the system, therefore analysis presented holds.

Let $x_i^+$ and $y_i^+$ denote the next state and output in time of state $x_i$ respectively,
\begin{eqnarray*}
    Lx_i &=& \tilde{L}ux_i\xi\\
    LE_{d\xi}E_{du}x_i\xi &=& \tilde{L}ux_i\xi
\end{eqnarray*}
Therefore, $BN$ is  equivalent to $BCN$ in state transition, respectively output sequence, if $L_D = \tilde{L}$, respectively if $HL_D = H\tilde{L}$, where $L_D := LE_{d\xi}E_{du}$ for all inputs. i.e. for every input, disturbances are effectively benign. Therefore the disturbance is decoupled from $BCN$ implying that $\tilde{L}$ can be divided into blocks of rank one.

For state feedback, the $BN$-$BCN$ equivalence output sequence requires $OS(Col_iL)=OS(Col_i\tilde{L}_x)$ and state transition equivalence requires $Col_i(L)=Col(Blk_i\tilde{L}_x)$. For output feedback, the $BN$-$BCN$ equivalence output sequence requires $OS(Col_iL)=OS(Col_i\tilde{L}_y)$ and state transition equivalence requires $Col_i(L)=Col(Blk_i\tilde{L}_y)$ where, $\tilde{L}_x$ and $\tilde{L}_y$ are the state transition matrices formed by using $M_x$ and $M_y$ as state and output feedback structures respectively, $Blk_i(.)$ indicates set of column vectors of the $i^{th}$ block of a matrix.

\textbf{Application:}
Available methods of DD are restrictive in the sense that the trajectory of system evolution is forced to a path that might not be natural. Such restriction on system evolution in critical systems like biological systems may be too imposing, even damaging to the system. Generally, with such systems the aim is to follow the natural course and nullify the effect of disturbance with external input/stimulus. Control signal or controller is not expected to alter the system trajectory. Control efforts should be focused on counteracting the uncertainties. In case of such requirement it is useful to be able to analyze for possible system models that satisfy the essential system behaviour criterion to identify the best suited model with least restriction.
    
The analysis presented here provides the characterization of the systems where the above mentioned goal is achievable.
\begin{figure}[t]
\centering
\includegraphics[scale=1]{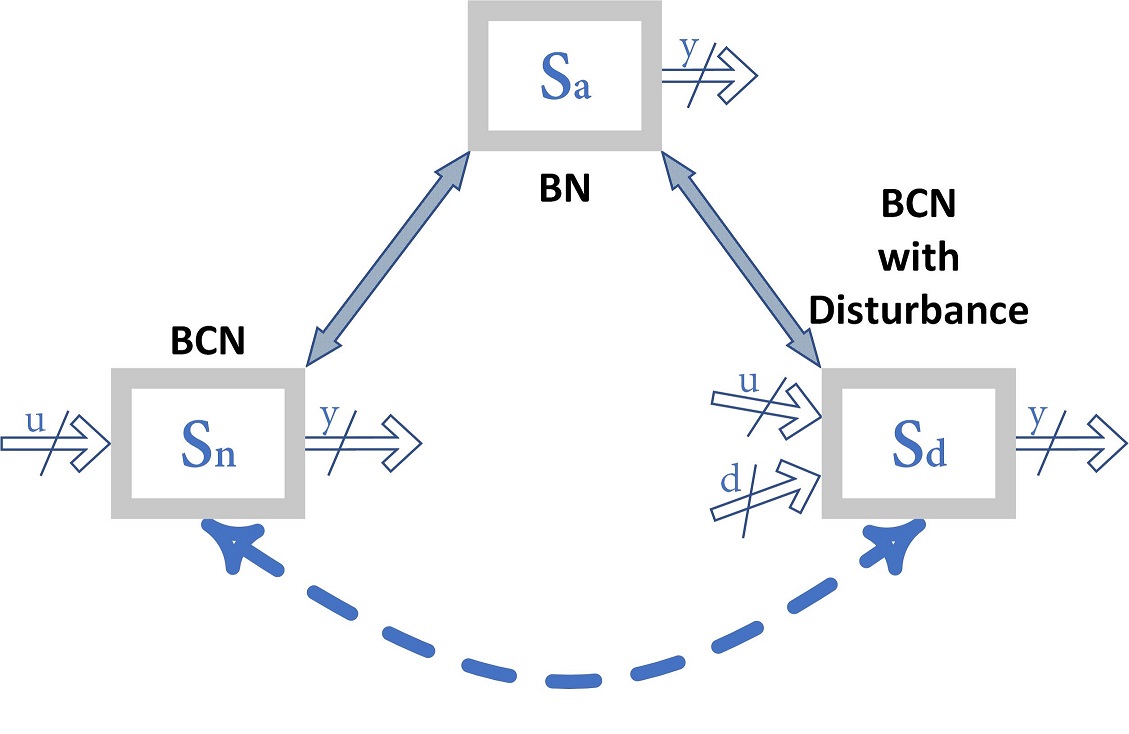}
\caption{Equivalence in a Nutshell}
\label{Equivalence_in_a_Nutshell}
\end{figure}

\textbf{Eg. Number of equivalent Boolean Systems}\\
The total number of possible $BN$ of $n$ variables is $(f_t^n):=2^{2^n}$. A Boolean function for each variable can be selected from $f_t^n$ available. Therefore the total number of possible $BNs$ in $n$ variables is 
\begin{equation*}
    T_v^n:=\underset{n\  times}{\underbrace{2^{2^n}\times 2^{2^n}\times \dots \times 2^{2^n}}}
    \   i.e. \   T_v^n:=2^{n.2^n} 
\end{equation*}

This result can be alternatively obtained as,\\
Total number of states (canonical vectors) possible for $n$ variables $S_t:=2^n$. Every state can transition to any one of $2^n$ states in possible $BNs$. Therefore, the number of possibilities of $BN$ is
\begin{equation*}
    T_s^{2^n}:=\underset{n\  times}{\underbrace{2^{n}\times 2^{n}\times \dots \times 2^{n}}}
    \   i.e. \   T_s^{2^n}:=2^{n.2^n} 
\end{equation*}
It can be seen that $T_v^n=T_s^{2^n}$.

The latter approach can be generalized to any network with $m$ number of nodes as $T_s^m=m^m$, where $m$ is not necessarily a power of $2$ i.e. $2^k=m$. This is useful in computing the total number of sub-network possibilities when the network structure is partially fixed. 

Suppose there are a total of $S_t$ states (nodes) of a $BN$, of which $S_c$ are a part of one of the cycles $C$ that executes the desired behaviour. The remaining $S_r:=S_t - S_c$ states transition to some state from $S_c$. Then the total number of $BNs$ that posses this structure can be calculated in a manner described below.

Consider the entire set $S_c$ to be a single state (node) of the network. Then the total possible $BNs$ with this modified structure, $N_{mod}:=(S_r+1)^{(S_r+1)}$. It can be verified that out $N_{mod}$ possible $BNs$, in only $N_{mod}/(S_r+1)$ $BNs$ the $S_c$ state transitions to itself (node $S_c$ has a self-loop) and hence remains invariant. \footnote{This can be observed by listing all the possible transitions from every node similar to a truth table, with $S_c$ set as a single node and all the remaining nodes. Since all the possibilities are listed as $S_c$ must transition to every node equal number of times and since there are a total of $(S_r+1)$ nodes, the number of possible transitions to every node is the total number of possibilities $((S_r+1)^{(S_r+1)})$ divided by $(S_r+1)$ i.e. $\displaystyle  \frac{(S_r+1)^{(S_r+1)}}{(S_r+1)}$. \textbf{E.g.} Draw the table similar to the truth table. Then every row represents one possible network structure, where the entries in the row corresponding to each column indicate transition from the header entry node to the row entry node. Consider the following case with $S_r$ containing two nodes say $N_1$ and $N_2$ respectively. Then the all possible network structures with self edge on $S_c$ can be represented as 
$$\begin{matrix}
\mathbf{S_c} & \mathbf{N_1} & \mathbf{N_2}\\ 
S_c & N_1 & N_1\\ 
S_c & N_1 & N_2\\ 
S_c & N_1 & S_c\\ 
S_c & N_2 & N_1\\ 
S_c & N_2 & N_2\\ 
S_c & N_2 & S_c\\ 
S_c & S_c & N_1\\ 
S_c & S_c & N_2\\ 
S_c & S_c & S_c
\end{matrix}$$
which is a third of the total possible network structures $3^3=27$.}

Therefore, the number of invariant possibilities is $    N_{mod}^{inv} = \displaystyle \frac{N_{mod}}{S_r+1}$. Since the interest here is to determine the number of possible networks that mimic the partial behaviour of a sub-network, which is also assumed to be invariant, only 
the self-transition is needed to be considered. These $N_{mod}^{inv}$ possibilities will also include the systems with $S_c$ as an unreachable set. In other words, the network graph of the modified network will be disconnected for these systems with number of disconnected components varying between $2$ to $(S_r+1)$. Of-course this is possible only if the remaining nodes contain fixed point/s and/or cycle/s. The fixed points will be characterised by the self-loop and cycles by cyclic directed paths i.e. loops of length 2 or more. 

By construction of the transition table it is clear that, every node appears equal number of times including the node corresponding to the column itself resulting in the self loops. Every column has $(S_r+1)$ such elements, distributed evenly among all the possibilities of the remaining columns, implying that every column has $\frac{1}{S_r+1}$ components, rows corresponding to which are needed to be discarded. Therefore, every column will reject $\left(\frac{1}{S_r+1}\right)^{th}$ of the available possibilities and allow $\left(\frac{S_r}{S_r+1}\right)^{th}$ of them for further scrutiny. Denoting the remaining possibilities after the removal of the self-looping elements by $N_1$,
\begin{eqnarray*}
    N_1 &=& N_{mod}^{inv} \times \underset{S_r\ times}{\underbrace{\frac{S_r}{S_r+1} \times \dots \times \frac{S_r}{S_r+1}}}\\
        &=& \frac{(S_r+1)^{(S_r+1)}}{(S_r+1)}\times \frac{(S_r)^{S_r}}{(S_r+1)^{S_r}}\\
    N_1 &=& S_r^{S_r}    
\end{eqnarray*}
For the removal of the possibilities pertaining to cycles from $N_1$, all the loops of length $2$ to $S_r$ need to be considered; which can be accomplished with the help of various possible permutations and combinations as follows:

All the possible loops of length $n$ say $n$-loop, can be built in three steps
\begin{enumerate}
    \item Selecting n nodes out of $S_r$ for the loop in $_{n}^{S_r}\textrm{C}$ ways
    \item Constructing n-loop from selected nodes. Two important characteristics of loops/ cycles are useful in this construction
    \begin{itemize}
        \item Self-loops are not possible at any node in any cycle of length $2$ or more. Therefore the same index as that of the node can not be used for transition. This reduces all the possibilities available at every node by one.
        \item Any n-loop can not have m-loop, with $m < n$, inscribed in it. Therefore, any subset of indices creates a loop smaller than n in length is not allowed.
    \end{itemize}
    This creates a nice structure for all possible n-loops as:
    \begin{center}
        \begin{tabular}{ c c c c c c c c}
        index & 1 & 2 & . & . & . & (n-1) & n \\ 
        possibilities & (n-1) & (n-2) &\ &\ &\ & 1 & 1
        \end{tabular}
    \end{center}
    Therefore, the total number of n-loops possible are,
    \begin{eqnarray*}
        N_l^n &=& (n-1).(n-2)\dots 2.1.1\\
              &=& (n-1)!
    \end{eqnarray*}
    \item Selecting the remaining $(S_r-n)$ nodes in $S_r^{(S_r-n)}$ ways (Every node can be chosen from $S_r$ possibilities excluding possibilities of self-loop already covered in $N_1$). The total n-loops possible    $\  N_n =\ ^{S_r}\textrm{C}_n\cdot (n-1)!\cdot S_r^{(S_r-n)}$.
\end{enumerate}
 The total number of connected modified sub-networks possible is at least $N_{mod}^c:= N_1 - \sum_{i=2}^{S_r} N_i $. Note that the phrase `at least' is used to indicate the fact that this is a conservative estimate, in the sense that every possible n-loop is included in the calculation separately, without considering the possibilities of systems with multiple loops (generated by $(S_r-n)$ remaining nodes) contributing to the count more than ones. These possible networks of remaining states can then be connected to any one of the $S_c$ states (nodes) of the fixed sub-networks with the same behaviour (in steady state) as that of sub-network $S_c$ with $N_T^{S_c}=\left | S_c \right |\cdot N_{mod}^c=S_c\cdot N_{mod}^c$.
 
 As one can observe, the assumption of invariance of the $S_c$ is not vital for the line of reasoning followed and consequently, the analysis can be extended for other cases. The primary aim of the analysis was to estimate the number of structurally similar $BN$, and it shows that the number grows faster than exponents (or exponentially) with $\left | S_c \right |$ and $\left | S_r \right |$.


\section{Output Feedback Stabilization}
The stabilization of the system states can be categorized into two types if the following conditions based on directed network graph are satisfied:
\begin{enumerate}
    \item To any state $x\in \delta_{2^n}$
        \begin{itemize}
            \item Self loop on $x$ i.e. $Col(\tilde{L})=x$
            \item Directed path exists from any state to $x$
        \end{itemize}
    \item To any set $X_s\subset \delta_{2^n}$
        \begin{itemize}
            \item $X_s$ is strongly connected in the network digraph
            \item Directed path exists from every state to $X_s$
        \end{itemize}
\end{enumerate}
\subsection{Output Feedback:}
Let the system dynamics be defined as $x^+=Lux$ and $y=Hx$, utilizing $u$ as the output feedback $u:=M_yy=M_yHx$ the system dynamics can be rewritten as
\begin{eqnarray}
    x^+=LM_yH\psi_nx
\end{eqnarray}
and 
\begin{multline}
    \tilde{H}:= H\psi_n = [Col_1(H)\otimes Col_1(I_{2^n})|\ Col_2(H)\otimes \\ Col_2(I_{2^n})\dots Col_{2^n}(H)\otimes Col_{2^n}(I_{2^n})]
\end{multline}
where 
\begin{multline}
    \tilde{H}_F:= M_y\tilde{H} = [Col_{Col_1(\tilde{H})}(M_y\otimes I_{2^n}) |\\ Col_{Col_2(\tilde{H})}(M_y\otimes I_{2^n}) \dots Col_{Col_{2^n}(\tilde{H})}(M_y\otimes I_{2^n})]
\end{multline}
Defining $\tilde{L}:=LM_yH\psi_n$ it can be expressed as
\begin{multline}
    \tilde{L}=[Col_{Col_1(\tilde{H}_F)}(L) | Col_{Col_2(\tilde{H}_F)}(L) \dots\\ \dots Col_{Col_{2^n}(\tilde{H}_F)}(L)]
\end{multline}
It can be observed that for $\tilde{L}$ its column set is a subset of the column set of the $L$ i.e. $Col(\tilde{L})\subseteq Col(L)$ and the columns that appear in the $\tilde{L}$ are decided by the feedback matrix $M_y$.

An algorithm is presented for the output feedback stabilization of a $BCN$:\\
\begin{algorithm}
\caption{Output Feedback Stabilization}\label{alg:OP_FB_Stab}

\begin{enumerate}
    \item Identify the column/s required, $C_{ri}$, at any index $i\in \{1,\dots ,2^n\}$.
    \item Check in $L$ if the required column/s exist(s).
    \item If $C_{ri}\notin Col[L]$, then the requirement can not be full filled $\Rightarrow$ break
    \item If $C_{ri}\in Col[L]$, then identify the index/ indices of the required column/s in $L$.\\
    i.e. $Column\_index\_set\_required(CS_i)$ defined as $CS_i:=\{j|C_{ri}=Col_j(L)\}$
    (requirement: $Col_j(L)=Col_i(\tilde{L})$ for some $i\ \& \ j$)
    \item (To see, if the requirement can be full filled)\\
    Check if, $\exists C_a$ such that $C_a\in Col(\tilde{H}_F)$ and $C_a\in CS_i$ i.e. $C_a\in Col(\tilde{H}_F)\bigcup CS_i$\\
    If NO $\Rightarrow$ Break\\
    If YES $\Rightarrow$ $Col_i(\tilde{L})=Col_{Col_i(\tilde{H}_F)}(L)$
    (requirement: $Col_i(\tilde{H}_F)\in CS_i$)
    \item $Col_i(\tilde{H}_F)=Col_{Col_i(\tilde{H})}(M_y\otimes I_{2^n})$, therefore to satisfy the requirement
    $$Col_{Col_i(\tilde{H})}(M_y\otimes I_{2^n})\in CS_i$$
    Check if $\exists\ C_b\in Col(M_y\otimes I_{2^n})\bigcup CS_i$\\
    If \textbf{NO} $\Rightarrow$ Break\\
    \textbf{Else} $\Rightarrow$ check if, $Col_{Col_i(\tilde{H})}(M_y\otimes I_{2^n})\in CS_i$\\
\end{enumerate}

\end{algorithm}

Next a few useful results are presented linking paths in the directed graph with feedback control of the system.

\textbf{Lemma 5:} Under any fixed control law, repeated node/s in a  path implies cycle.

\textbf{Corollary 6:} In any non-cyclic path, if nodes are repeated then the underlying control law is not fixed.\\
It may indicate changing control law, which is not the type of the control behaviour assumed in this work.

\textbf{Proposition 7:} While searching for directed paths the nodes should not be repeated, except for the cycles, where only the starting node also appears as the end node.

In terms of the conventional control the output feedback requirement can be stated as: all the states corresponding to the same output group (group of states for which the output is the same) should translate to the same input. In other words, in combined digraph (for every input) only the directed paths that represent the same control input for the states belonging to same output set, are useful for the output feedback. Based on this a procedure is suggested for the output feedback stabilization:
\begin{enumerate}
    \item Find all paths that satisfy the proposition.
    \item For every set of paths (every node appears only ones) identify the $(state-input)$ tuple for all [nodes-edges].
    \item List out the paths that follow the output feedback requirement.
    \item If any such path exists then the output feedback matrix is given by the corresponding $(state-input)$ tuple as,
    $$Col_{state\_output\_set}(M_y)=input$$
\end{enumerate}

\textbf{E.g. 1.}
\begin{eqnarray*}
    L'= & \delta_4 [\ 2\ 3\ 4\ 4\ 6\ 7\ 8\ 4\ 1\ 4\ 3\ 5\ 4\ 2\ 3\ 3\\ 
    & \ \; \; 1\ 1\ 3\ 4\ 5\ 2\ 7\ 8\ 3\ 3\ 4\ 4\ 5\ 5\ 7\ 7\ ] 
\end{eqnarray*}
$$H=\delta_4[1\ 1\ 2\ 2\ 3\ 3\ 4\ 4]$$
for $M_y=\delta_4[1\ 3\ 4\ 2]$ the resulting state transition matrix is given by
$$\tilde{L}:=\delta_8[2\ 3\ 3\ 4\ 5\ 5\ 3\ 3]$$
\textbf{E.g. 2.}
$$L=\delta_8[2\ 3\ 4\ 4\ 6\ 7\ 8\ 4\ 1\ 4\ 3\ 5\ 4\ 2\ 3\ 3]$$
$$H=\delta_4[1\ 2\ 1\ 2\ 1\ 2\ 1\ 2]$$
for $M_y=\delta_2[2\ 1]$ the resulting state transition matrix is given by
$$\tilde{L}:=\delta_8[1\ 3\ 3\ 4\ 4\ 7\ 3\ 4]$$

\textbf{Method:}\\
For states $x_i\in \delta_{2^n}$ corresponding to output $y_j\in \delta_{2^p}$; if $Col_{y_j}(M_y)=u_k\in \delta_{2^m}$, then for those states $Col_{x_i}(\tilde{L})=Col_{x_i}(L_{u_k})$ \footnote{Total number of possible output feedback functions: $2^{2^p}$, for 2 outputs: $2^{2^2}=16 \Rightarrow 4$ of them constant functions. Constant function $\rightarrow$ pinning control, e.g. $M_y=\delta_4[1\ 1\ 1\ 1\ 1\ 1\ 1\ 1]$}

\textbf{E.g. 3.} If
$$\begin{matrix}
L= & \delta_8 [\ 2\ 3\ 4\ 4\ 6\ 7\ 8\ 4\ 1\ 4\ 3\ 5\ 4\ 2\ 3\ 3\\ 
    & \; 1\ 1\ 3\ 4\ 5\ 2\ 7\ 8\ 3\ 3\ 4\ 4\ 5\ 5\ 7\ 7\ ]
\end{matrix}$$
$$H=\delta_4[1\ 3\ 4\ 1\ 2\ 3\ 1\ 4]$$
then there does not exist any $M_y$ that results in 
$$\tilde{L}=\delta_8[3\ 3\ 3\ 5\ 6\ 7\ 3\ 3]$$


\section{Disturbance Decoupling}
Considering the literature available, to eliminate disturbance from the system, the rank condition \cite{yang2013controller} needs to be satisfied for every block of the state-transition matrix that designates to disturbances and the states not appearing in the output equation.

\cite{yang2013controller} provided an algorithm for DD, which checks for possibility of DD and also provides possible state feedback laws.

But there are a few limitations with this approach:
\begin{itemize}
    \item The actual nature of output equation is not considered.
    \item Definition for $DDP$ and conditions for its solution (existence) are conservative and deviate from classical definition.
\end{itemize}
Consider the $BCN$ defined by
\begin{eqnarray}
    X^+=Lux\xi \\
    X^+=LM_x \psi_n x\xi
\end{eqnarray}
where the input is provided as state feedback, characterized by the matrix $M_x$. With $n-$ states and $d-$ disturbances the term $L\ltimes M_x \ltimes \psi_n=:L_x$ is equivalent to
\begin{align}\small
    L_x &= L\cdot (M_x \ltimes \psi_n \otimes I_{2^d}) \nonumber \\
    &= L \cdot \{[Col_1(M_x)\otimes Col_1(I_{2^n}) \   Col_2(M_x)\otimes Col_2(I_{2^n}) \nonumber\\ &~~~\dots Col_{2^n}(M_x)\otimes Col_{2^n}(I_{2^n})]\otimes I_{2^d}\}\nonumber\\
    &= L \cdot [Col_1(M_x)\otimes Col_1(I_{2^n}) \otimes I_{2^d}\ \nonumber \\ &~~~\dots Col_{2^n}(M_x)\otimes Col_{2^n}(I_{2^n}) \otimes I_{2^d}]\nonumber\\
    &= L \cdot \{[Col_1(M_x)\otimes Col(1:2^d|\delta_2^{n+d})]\nonumber\\ &~~~[Col_2(M_x)\otimes Col(2^d+1:2\cdot2^d|\delta_2^{n+d})]\nonumber\\ &~~~\cdots [Col_{2^n}(M_x)\otimes Col((2^n-1)2^d+1:2^{n+d}|\delta_2^{n+d})] \}
\end{align}
i.e. choosing/selecting $2^{n+d}$ columns of $L$ using $M_x \psi_n I_{2^d}$ where 
\begin{multline*}
M_x \psi_n \otimes I_{2^d}=[2^d\  columns\  of\  \delta_{2^{m+n+d}}^i| a\leq i\leq b] \\ such\  that\  \delta_{2^{m+n+d}}^a=\delta_{2^{n+d}}^{2^{[(2^{k-1}-1)2^d+1]}}\otimes \delta_{2^m}^1\ \\ and\  \delta_{2^{m+n+d}}^b=\delta_{2^{n+d}}^{2^{(k2^d)}}\otimes \delta_{2^m}^{2^m} \ for\  k\in \{1,2,\dots , 2^n\}
\end{multline*}
Each block picks $2^d$ consecutive columns of $L$, indexed between $a$ and $b$ only. Therefore possibility of disturbance decoupling and fault detection is solely affected by the transition matrix $L$.

The available necessary and sufficient conditions can in general be summarized as follows:

For a Boolean control network affected by the disturbances, to effectively decouple the disturbance, its state transition matrix $L$ needs to satisfy following criteria
\begin{itemize}
    \item If $\tilde{L}$ represents the state transition matrix of the output subsystem, also referred to as output friendly subspace, then divide $\tilde{L}$ into $2^m$ blocks of size $(2^r\times 2^{n+d})$ indicating by $\tilde{L}_{u_i}\forall i\in \{1, \dots , 2^m\}$
    \item Divide every $\tilde{L}_{u_i}$ into $2^s$ blocks of size $(2^r\times 2^{n-s+d})$ denoted by $\tilde{L}_{u_i}\forall j\in \{ 1,\dots , 2^s \}$ 
    \item Disturbance can be decoupled if and only if corresponding to every $X_j$ for $j\in \{ 1, \dots , 2^m \}$ $\exists i\in \{1, \dots , 2^m\}$ such that $f(\tilde{L}_{u_i}^{x_j})=1$ i.e. $\tilde{L}_{u_i}^{x_j}$ follows rank condition.
\end{itemize}
Denoting by $X_{1-s}$ set of all the states of the output subsystem. For disturbance to be decoupled, in general $\forall x_s \in X_{1-s}$ an input should exist such that $[\tilde{L}_{u_i}^{x_s}]$ corresponding to that input has rank 1.

One obvious issue with this approach is that, as presented in the sequel, it leaves out a large number of systems for which the disturbance can be decoupled with satisfactory performance.

\textbf{E.g. 4.} Let a system be defined by
\begin{eqnarray*}
    x_s^+ = \tilde{L}ux_1x_2x_3\xi\\
    y = \delta_2[2\ 1\ 1\ 2]x_1x_2
\end{eqnarray*}
with desired behaviour of the system is to track the reference $r=1$, when present. It is clear from the output equation that to track the reference $r=1, y=1$ i.e. the subsystem state $x_s\in \{x_1\bar{x_2}, \bar{x_1}x_2\}$ or $\{\delta_4^2,\delta_4^3\}$. If the system dynamics are such that the resulting $\tilde{L}$ matrix has the form, 
\begin{eqnarray*}
    \tilde{L}_{u_i} = \delta_4[e^1 e^2 e^3 e^4 | e^5 e^6 e^7 e^8 | e^9 e^{10} e^{11} e^{12} | e^{13} e^{14} e^{15} e^{16}]
\end{eqnarray*}
for any $i\in \{1,2\}$, with 
\begin{eqnarray*}
    e^j\in \delta_4\{2,3\}\    \forall j\in \{5,6,7,8,9,10,11,12\}
\end{eqnarray*}
OR
\begin{eqnarray*}
    e^j\in \delta_4\{2,3\}\    \forall j\in \{1,2,3,4,13,14,15,16\}
\end{eqnarray*}
OR
\begin{eqnarray*}
    e^j\in \delta_4\{2,3,4\}\    \forall j\in \{1,2,3,4\}\\
    and\\
    e^j\in \delta_4\{2,3\}\    \forall j\in \{13,14,15,16\}
\end{eqnarray*}
OR
\begin{eqnarray*}
    e^j\in \delta_4\{2,3\}\    \forall j\in \{1,2,3,4\}\\
    and\\
    e^j\in \delta_4\{1,2,3\}\    \forall j\in \{13,14,15,16\}
\end{eqnarray*}
For instance; let $\tilde{L}_1=\delta_4[2\ 3\ 2\ 3|3\ 3\ 3\ 3|2\ 2\ 2\ 2|1\ 2\ 3\ 3]$ for $u=1$ without loss of generality. Applying the state feedback law $M_x=\delta_4[1\ 1\ 1\ 1\ 1\ 1\ 1\ 1]$, when the reference is $r=1$, will stabilize the output at $y=1$ in at-most 2 evolutions of the dynamics irrespective of the initial state and the structure of $\tilde{L}_2$, where $\tilde{L}=[\tilde{L}_1|\tilde{L}_2]$. This obviously decouples the effect of disturbance from the output, although a large number of possible $\tilde{L}$ matrices do not satisfy the required conditions.

Additionally, if $e^i\in \delta_4\{2,3\}\ \forall i\in \{1,2,3,4,13,14,15,16\}$ i.e. say $\tilde{L}_1=\delta_4[2\ 3\ 2\ 3|3\ 3\ 3\ 3|2\ 2\ 2\ 2|2\ 2\ 3\ 3]$ without loss of generality; then the performance of system will be as good as any system following the necessary and sufficient condition. This example supports the claim that the available necessary and sufficient conditions are conservative.

To develop more inclusive necessary and sufficient conditions, the ideas expanding the current notions of reachability and disturbance decoupling are discussed next. 
\subsection{Reachability}
Reachability in its regular sense indicates all the points in the state space that can be reached starting from any point following the system dynamics. Any point $a$ is said to be reachable from any point $b$ if under system dynamics and appropriate input, starting from $b$ the system trajectory reaches point $a$ in finite time. In Boolean domain the concept of reachability is defined on the same lines. Three notions of reachability are presented here to aid the analysis of disturbance decoupling.\footnote{ \cite{li2013state} defined $E_k(r)$ as the set consisting of all the initial states that can be steered to $\delta_{2^n}(r)$ in $k$ steps by some control input sequence. Clean reachability and definite reachability satisfy this definition. Indefinite reachability, however, can be regarded as a diversified approach established to accommodate uncertainty in state transition.}

\subsubsection{Clean Reachability}
State $a$ is said to be cleanly reachable from state $b$ if for some input under the system dynamics, state $b$ transitions to state $a$ directly. Denoting the same by $b\overset{C}{\rightarrow}a$.

i.e. $b\overset{C}{\rightarrow}a$ iff $\exists i\in \{1,\dots , 2^m\}$ such that $Lu_ib=a$, where $L:$ the state transition matrix, $u_i:$ $i^{th}$ input belonging $\delta_{2^m}$ and $a,b:$ system states i.e. $a,b\in \delta_{2^n}$

\subsubsection{Definite Reachability}
State $a$ is said to be definitely reachable from state $b$ if for some input sequence under the system dynamics, starting from state $b$ the system trajectory reaches state $a$ in a finite (predefined) number of evolutions. Denoting the same by $b\overset{d}{\rightarrow}a$, 

$b\overset{d}{\rightarrow}a$ iff $\exists i_1,i_2,\dots ,i_k\in \{1,\dots , 2^m\}$ such that $Lu_{i_1}u_{i_2}\dots u_{i_k}b=a$, where $k$ is a positive integer known in advance.

\subsubsection{Indefinite Reachability}
State $a$ is said to be indefinitely reachable from state $b$ if for some input sequence of unknown length under the system dynamics, starting from state $b$ the system trajectory may eventually reach state $a$. Denoting the same by $b\overset{id}{\rightarrow}a$,

$b\overset{id}{\rightarrow}a$ if $\exists i_1,i_2,\dots ,i_k\in \{1,\dots , 2^m\}$ such that $Lu_{i_1}u_{i_2}\dots u_{i_k}b=a$, where $k$ is some positive integer.
\subsection{Disturbance Decoupling}
As is the case with the reachability, the notion of disturbance decoupling in $BCN$ can also be categorized. Three notions of DD are defined here, which are helpful for the further work.

\subsubsection{DD in Mapping}
The $BCN$ can be represented as a $(state,input)\mapsto (next\_state)$ and the effect of disturbance can be incorporated as uncertainties in the map. Therefore, DD in mapping implies removal of uncertainties in/from the map. A system is disturbance decoupled in its mapping if its $(X,I)\mapsto (n_X)$ map is deterministic (for output friendly space). The accepted/available notion of DD in the literature can be categorised as DD in mapping with completely deterministic map i.e. removal of uncertainties for all $(state, input)$ tuples.

\subsubsection{DD in Iterations}
The removal of uncertainties could be limited to certain $(state, input)$ tuples. If a set is constructed containing states from all such tuples such that this set contains an invariant set $(S_1)$ under some input and if all the remaining states definitely reach the invariant set. Then in this case DD is said to be achieved in iterations, as starting from any initial condition the effect of disturbance will vanish in at-most $k$ systems evolutions, where $k$ is the largest number of the system evolutions required to reach the invariant set.

\subsubsection{DD Invariant in Output (Y)}
In addition to DD in mapping (iteration) if the output sets $(O_s)$ are controlled invariant under the system dynamics, then it is defined as DD invariant in output $(Y)$ solvable, where $O_{Si}:=\{x_s\in X_{1-s}|H(x_s)=\delta_{2^p}^i\}$, i.e. set of all the sub-states of $X_{1-s}$ that have the same output.\footnote{System enters in a subset, which can be divided into output groups and the transition from one output group to another is unaffected by the disturbance.}

\subsection{Construction of Reachability Graphs}
As discussed earlier, there are three different notions of reachability. Establishing possibility for clean reachability is relatively straightforward. For definite and indefinite reachability, the process is slightly complicated. Directed graph comes as a handy tool to tackle this problem. To establish reachability to any output set $O_{sl}$ a graph is constructed as follows:

Let $\mathcal{V}:= \{ X_{1-s}\}=\{x_{si}|x_{si}\in X_{1-s}\}$ for $i \in \{1,2,\dots ,2^s\}$\\
\textbf{Definite:} $(a,b)\in \mathcal{E}$ if $\exists x_{sa}^i$ for some $i\in \{1,\dots, 2^m$ such that $x_{sa}^{i+}$ is a singleton set and $b\in x_{sa}^{i+}$ (i.e. edge is present iff under some control $a\xrightarrow[to]{transitions} b$ with certainty, where $a\ \&\ b$ are states $\in X_{1-s}$\\
\textbf{Indefinite:} $(a,b)\in \mathcal{E}$ if $\exists x_{sa}^i$ for some $i\in \{1,\dots, 2^m$ such that $b\in x_{sa}^{i+}$ (i.e. edge if possibility of transition $a\rightarrow b$).\\

\textbf{Proposition 8:} Any state $a\in X_{1-s}$ is cleanly reachable from any $b\in X_{1-s}$ iff there exists an edge from $b$ to $a$ in the graph constructed for definite reachability.

\textbf{Proposition 9:} Any state $a\in X_{1-s}$ is definitely reachable from any $b\in X_{1-s}$ iff there exists a path from $b$ to $a$ in the digraph constructed for definite reachability.

\textbf{Proposition 10:} Any state $a\in X_{1-s}$ is indefinitely reachable from any $b\in X_{1-s}$ iff there exists a path from $b$ to $a$ in the digraph constructed for indefinite reachability.

\textbf{Proof:} Construction of reachability digraphs in proposition 1 to 3 are based on reachability definitions, therefore proofs are trivial.$\square$

\begin{figure}[t]
\centering
\includegraphics[scale=0.7]{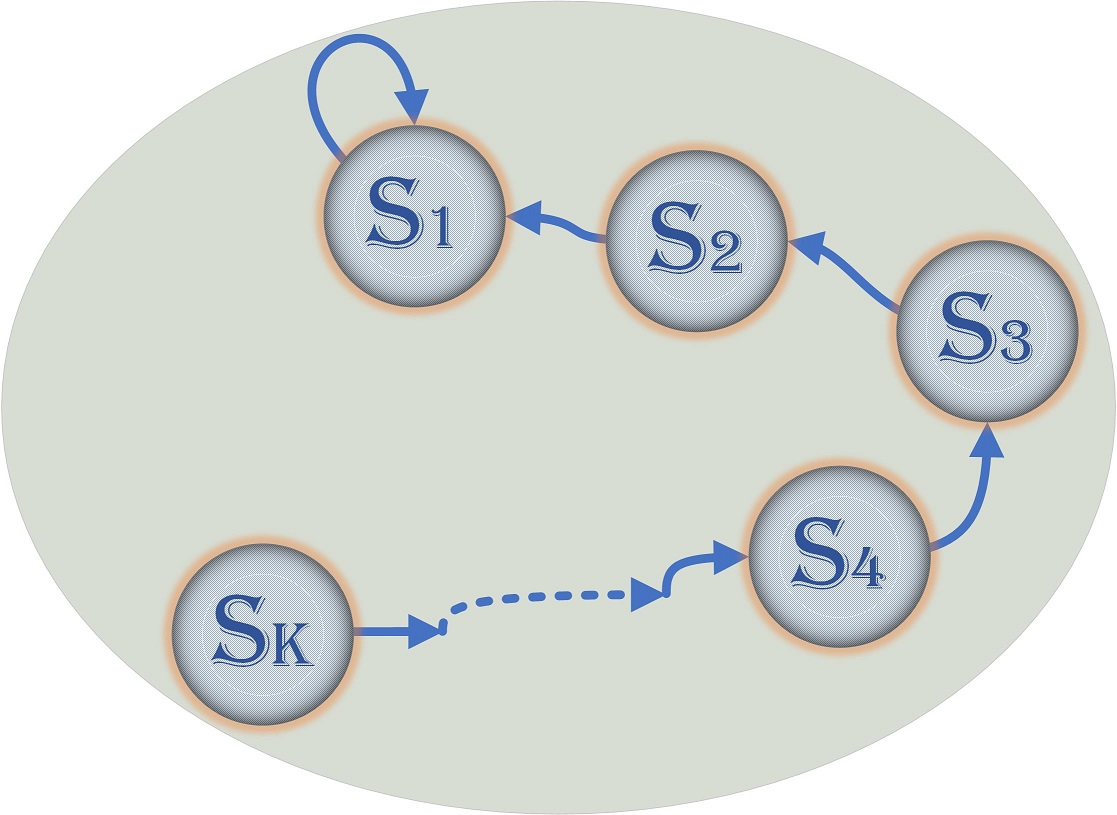}
\caption{Clean reachability of output sets}
\label{clean_output_reachability}
\end{figure}

\textbf{E.g. 5.} Suppose $f_y \ or\ H=\delta_2 [2\ 1\ 2\ 1]$ where $X\in X_{1-s}$, it is clear that $s=2$. Therefore $X_{1-s}=\{x_1x_2=\delta_4^1, x_1\bar{x}_2=\delta_4^2, \bar{x}_1x_2=\delta_4^3, \bar{x}_1\bar{x}_2=\delta_4^4$, for notational simplicity $X_{1-s}$ can be written as $X_{1-s}=\{X_1,X_2,X_3,X_4\}$. Let $O_{S1}$ correspond to $y_1=1=\delta_2^1$ and $O_{S2}$ correspond to $y_2=2=\delta_2^2$, therefore $O_{S1}=\{X_2,X_4\}$ and $O_{S2}=\{X_1,X_3\}$. Let $L=\delta_4[2\ 4\ 3\ 4|2\ 4\ 4\ 1|2\ 4\ 4\ 3|4\ 2\ 4\ 3]$, then digraph for output sets can be constructed as in Fig. \ref{clean_output_reachability}.

\begin{figure}[t]
\centering
\includegraphics[scale=0.55]{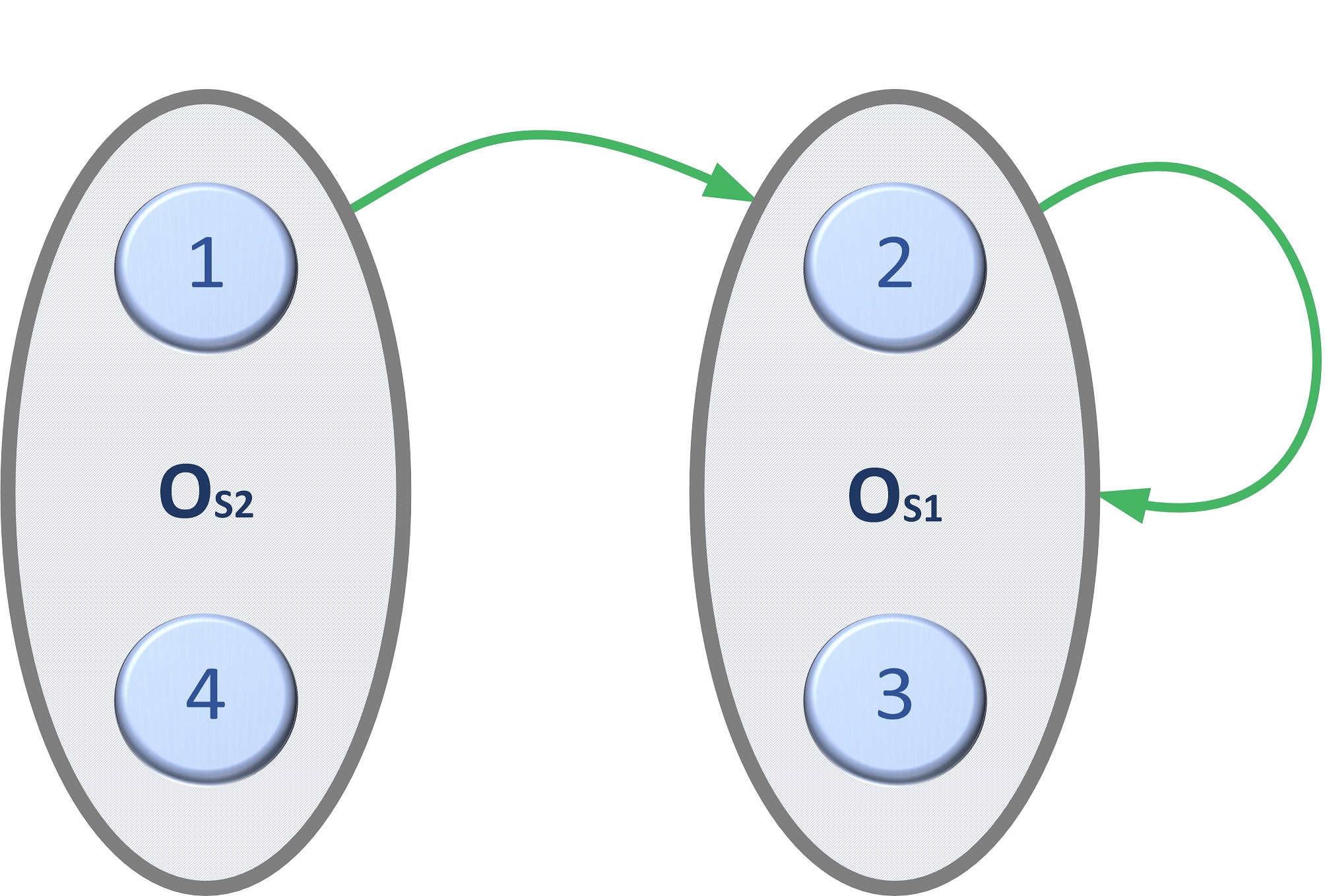}
\caption{Clean reachability of output sets}
\label{clean_reachability_output_sets}
\end{figure}

There is an edge from $O_{S2}$ to $O_{S1}$, because for both $X_1\ \& \ X_3$ control is available to transition with certainty to an element of $O_{S1}$. The same is not true for either of $X_2\ \& \ X_4$. Similarly, self-loop of $O_{S1}$ indicates transition certainty with its own element which is not the case with $O_{S2}$ as a result, no self-loop. Hence, $O_{S1}$ is both reachable and invariant. $O_{S1}$ here is cleanly reachable as can be seen by construction of $O_{k+}^i$ for $i\in \{1,2\},k\in \{1,2,3,4\}$.\\
$$\begin{matrix}
x_{sk}^{i+}:\Rightarrow  & X_1^{1+}=\{2,4\} & X_1^{2+}=\{3,4\}\\ 
 & X_2^{1+}=\{2,4\} & X_2^{2+}=\{1,4\}\\ 
 & X_3^{1+}=\{2,4\} & X_3^{2+}=\{3,4\}\\ 
 & X_4^{1+}=\{2,4\} & X_4^{2+}=\{3,4\}
\end{matrix}$$
$$\begin{matrix}
O_{k+}^{i}: & O_{1+}^1=\{1\} & O_{1+}^2=\{1,2\}\\ 
 & O_{2+}^1=\{1\} & O_{2+}^2=\{1,2\}\\ 
 & O_{3+}^1=\{1\} & O_{3+}^2=\{1,2\}\\ 
 & O_{4+}^1=\{1\} & O_{4+}^2=\{1,2\}
\end{matrix}$$

Reachability to $O_{S1}:y_1=1$\\
$X_1:1\in O_{1+}^1 \&\ O_{1+}^1$ is singleton, $\therefore C_{X_1}=\{1\}$ hence cleanly reachable\\
$X_2:1\in O_{2+}^1 \&\ O_{2+}^1$ is singleton, $\therefore C_{X_2}=\{1\}$ hence cleanly reachable\\
Similarly for $X_3\ \& \ X_4$ are singleton, therefore $O_{S1}$ is cleanly reachable.

Reachability to $O_{S2}:y_2=2$\\
$X_1:2\in O_{1+}^2 \ but\ O_{1+}^2$ is not singleton, $\therefore C_{X_1}=\{\phi \}$ hence, not cleanly reachable\\
Similarly, for $C_{X_i}=\{\phi \}$ for $i\in \{2,3,4\}$, therefore $O_{S2}$ is not cleanly reachable.

\textit{Definite Reachability:}
$\mathcal{V}:=\{X_1,X_2,X_3,X_4\}$

\begin{figure}[t]
\centering
\includegraphics[scale=0.55]{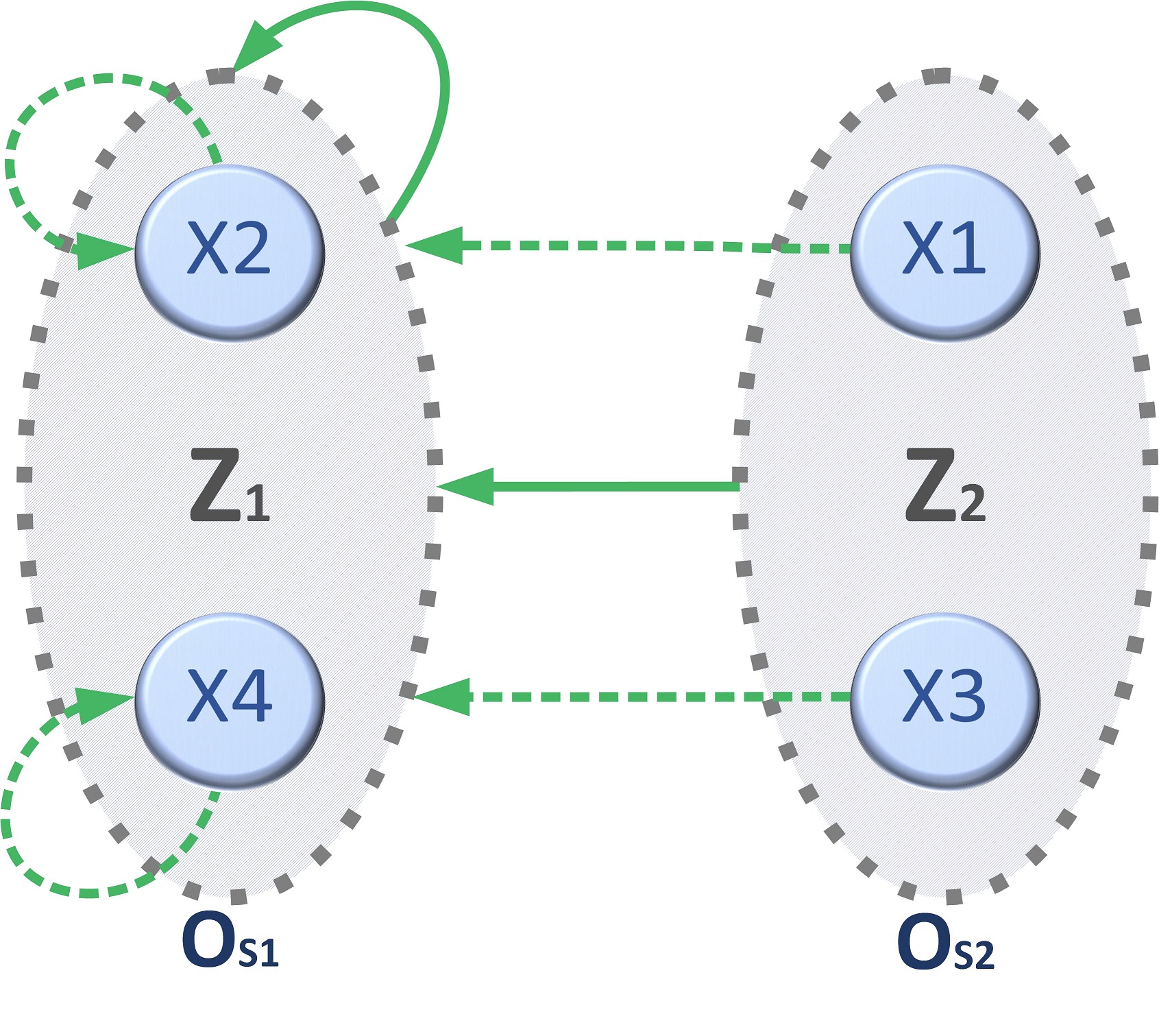}
\caption{Definite reachability of output sets}
\label{definite_output_reachability}
\end{figure}

\textbf{$O_{S1}$:} Since $O_{S1}$ is a set of vertices (states), a single large vertex is constructed by replacing all the vertices in the set by some other vertex, say $Z_1$, and carry out the test for definite reachability. Representing $X_2\ \& \ X_4$ by $Z_1$\\
\textbf{$X_1$}: $X_1^{1+}=\{X_2,X_4\}=\{Z_1,Z_1\}=\{Z_1\}\therefore$ definitely reachable from $X_1$\\
\textbf{$X_3$}: $X_3^{1+}=\{X_2,X_4\}=\{Z_1,Z_1\}=\{Z_1\}\therefore$ definitely reachable from $X_3$\\
Therefore, $O_{S1}$ is definitely reachable.

\textit{Indefinite Reachability:} The digraph is as:

\begin{figure}[t]
\centering
\includegraphics[scale=0.55]{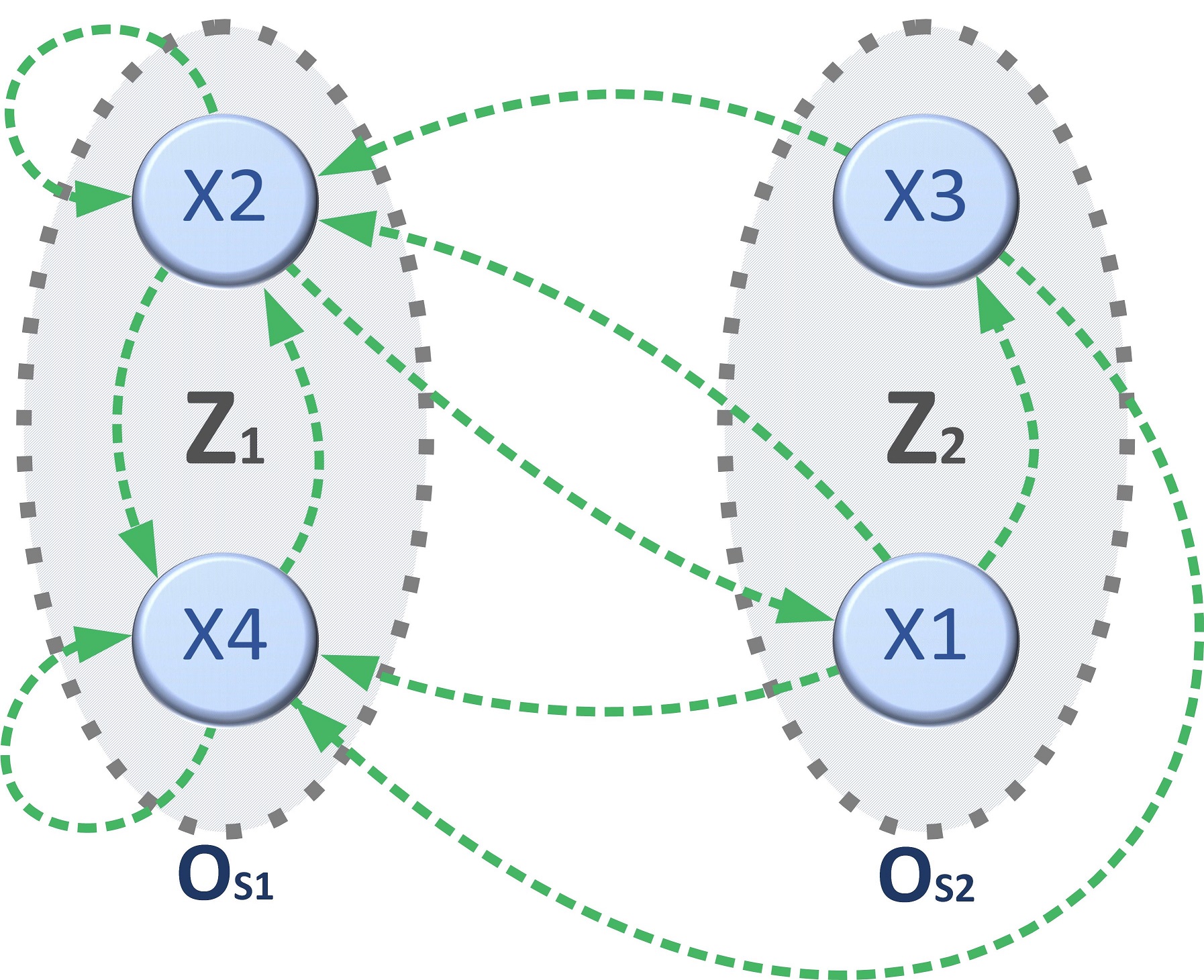}
\caption{Indefinite reachability of states}
\label{indefinite_state_reachability}
\end{figure}

It is easy to see that both $O_{S1}$ and $O_{S2}$ are indefinitely reachable, where $\mathcal{Z}_1=\{X_2,X_4\}$ and $\mathcal{Z}_2=\{X_1,X_3\}$.

An obvious observation one can make is that $Clean\  Reachability\Rightarrow Definite\ Reachability \Rightarrow Indefinite\ Reachability$. This can be easily explained by the fact that $Clean\  Reachability\ Conditions\Rightarrow Definite\ Reachability\ Conditions \Rightarrow Indefinite\ Reachability\ Conditions$. Since the inverse implication does not hold the converse is not true.

Depending upon the structure of the $L$ and $H$ matrices the feedback can be selected for every output friendly state to achieve different possibilities of DD and output reachability. E.g. a structure can be formed by subsequent priorities:
\begin{itemize}
    \item Make as many output sets invariant as possible
    \item Of the remaining output sets, make as many of them as possible to cleanly reach the invariant output set
    \item Make as many of the remaining output sets as possible to definitely reach one of the invariant sets
    \item Of the remaining output sets make as many as possible disturbance decoupled in mapping
    \item Of the remaining output sets as many as possible to indefinitely reach the invariant output set
\end{itemize}
These properties can be utilized in general for addressing any type of problem, owing to the fact that this adds an hierarchical structure to the system. The possibility of hierarchical structure makes this approach much more useful in controlling the behaviour of the system as illustrated in following sections. A system + controller form, analogous to classical controller-plant feedback, that responds to reference signal and eliminates disturbances emerges from this structure, which is not observed in the literature.

\textbf{E.g. 6.} \textit{DD: (In mapping)} let,\\
$\begin{matrix}
L'= & \delta_4[1\ 2\ 1\ 2\ |& 3\ 4\ 3\ 4\ |& 2\ 4\ 4\ 4\ |& 3\ 1\ 3\ 2\\ 
 & \ \ \ \ 2\ 3\ 2\ 2\ |& 4\ 2\ 4\ 2\ |& 1\ 3\ 4\ 1\ |& 3\ 3\ 3\ 1\\ 
 & \ \ \ \ 1\ 4\ 1\ 2\ |& 3\ 4\ 3\ 4\ |& 1\ 3\ 3\ 1\ |& 3\ 2\ 2\ 1\\ 
 & \ \ \ \ 2\ 4\ 2\ 2\ |& 4\ 1\ 4\ 4\ |& 1\ 3\ 3\ 2\ |& 4\ 4\ 4\ 2] 
\end{matrix}$
$$X_{12}=L'M_x \psi_nX_{123}\xi$$
$$f_y=\delta_2[2\ 1\ 2\ 1]$$
\begin{itemize}
    \item $y_1=\{2,4\}=O_{S1}\Rightarrow 1; \ y_2=\{1,3\}=O_{S2}\Rightarrow 2$
    \item $x_{sk}^{i+}:\ X_{1-s}={X_1=\delta_4^1, X_2=\delta_4^2, X_3=\delta_4^3, X_4=\delta_4^4}$\\
    $\begin{matrix}
    & i\rightarrow & 1=\delta_4^1 & 2=\delta_4^2 & 3=\delta_4^3 & 4=\delta_4^4\\ 
    X_1^{i+}= &  & \{\{1,2\} & \{2,3\} & \{1,2,4\} & \{2,4\}\}\\ 
    X_2^{i+}= &  & \{\{3,4\} & \{2,4\} & \{3,4\} & \{1,4\}\}\\ 
    X_3^{i+}= &  & \{\{2,4\} & \{1,3,4\} & \{1,3\} & \{1,2,3\}\}\\ 
    X_4^{i+}= &  & \{\{1,2,3\} & \{1,3\} & \{1,2,3\} & \{2,4\}\}
    \end{matrix}$
    \item $O_{k+}:$ for $k\in \{1,2,3,4\}/\{X_1,X_2,X_3,X_4\}$
    $\begin{matrix}
    & i\rightarrow & 1=\delta_4^1 & 2=\delta_4^2 & 3=\delta_4^3 & 4=\delta_4^4\\ 
    O_{X_1+}= &  & \{\{2,1\} & \{1,2\} & \{2,1\} & \{1\}\}\\ 
    O_{X_2+}= &  & \{\{2,1\} & \{1\} & \{2,1\} & \{2,1\}\}\\ 
    O_{X_3+}= &  & \{\{1\} & \{2,1\} & \{2\} & \{2,1\}\}\\ 
    O_{X_4+}= &  & \{\{2,1\} & \{2\} & \{2,1\} & \{1\}\}
    \end{matrix}$
\end{itemize}
Therefore, the possible controls for DD in mapping are:
$$M_x = \delta_4[4\ 4|2\ 2|1/3\ 1/3|2/4\ 2/4]$$
It can be observed that, only $O_{S1}$ can be made invariant and it is also reachable (cleanly). Therefore, under the action of control law
$$M_x = \delta_4[4\ 4\ 2\ 2\ 1\ 1\ 4\ 4]$$
the output of the system is stabilized to $H(O_{S1})=1=\delta_2^1$, whereas for remaining possible control laws the system is disturbance decoupled only in mapping.

\begin{figure}[t]
\centering
\includegraphics[scale=0.55]{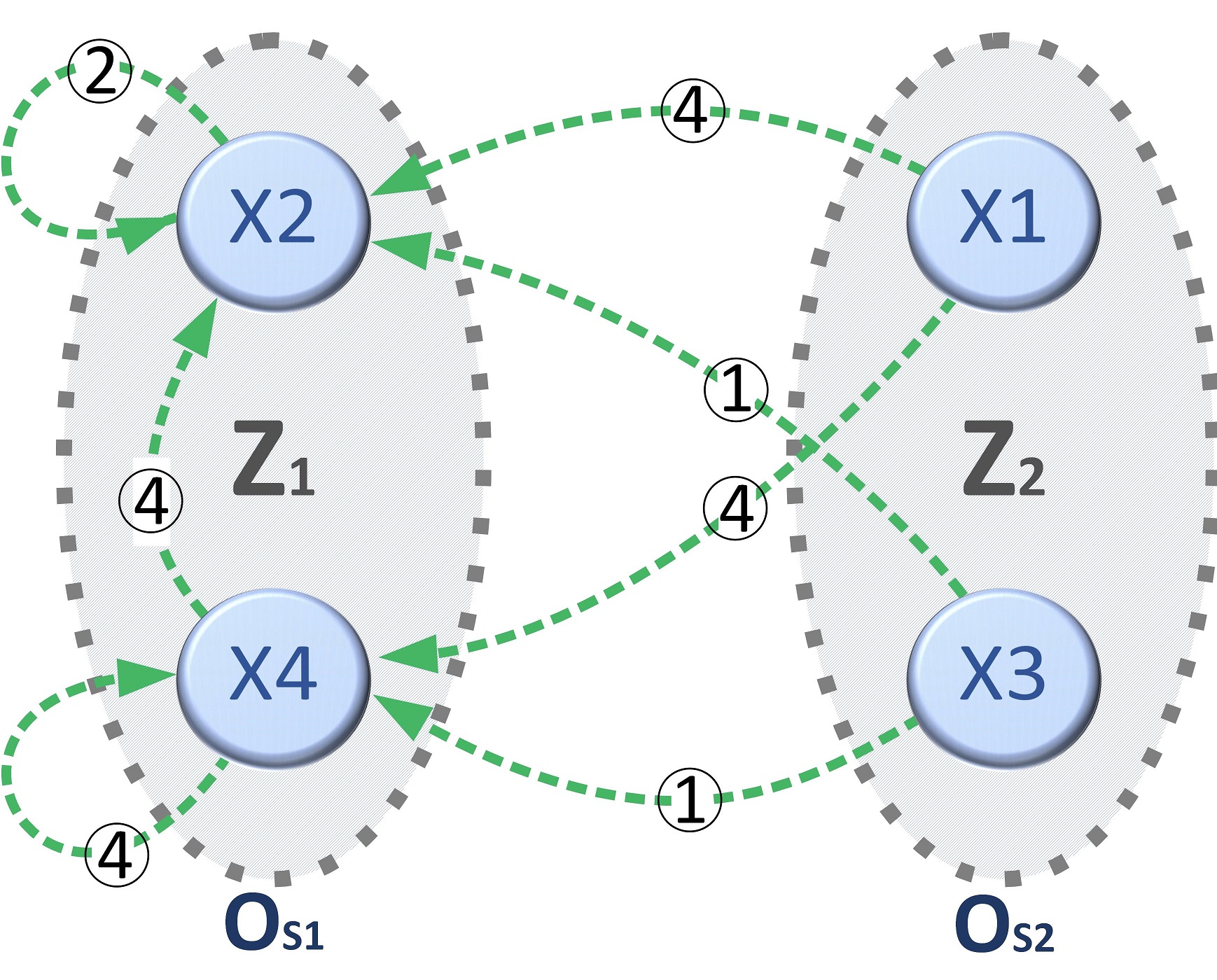}
\caption{Uncertain reachability of states with control}
\label{indefinite_state_reachability_wt_control}
\end{figure}

Control law for output stabilizing feedback in $O_{S1}$ or $Y=1$

\begin{figure}[t]
\centering
\includegraphics[scale=0.5]{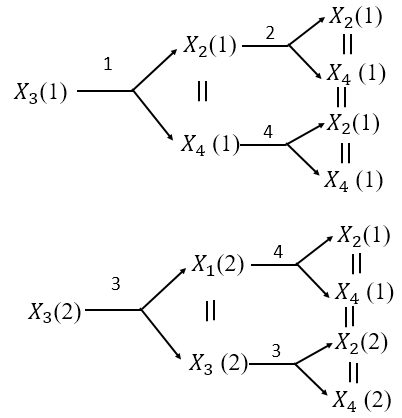}
\caption{Disturbance decoupling in mapping}
\label{DD_in_mapping}
\end{figure}
Therefore outputs, stabilizer:
$$\begin{matrix}
X_0: & X_1\Rightarrow 2\ 1\ 1\ \dots\\ 
 & X_2\Rightarrow 1\ 1\ 1\ \dots\\ 
 & X_3\Rightarrow 2\ 1\ 1\ \dots\\ 
 & X_4\Rightarrow 1\ 1\ 1\ \dots
\end{matrix}$$
Mapping:
$$\begin{matrix}
X_0: & X_1\Rightarrow & 2\ 1\ 1\ 1\ 1\ 1\dots\\ 
 & X_2\Rightarrow & 1\ 1\ 1\ 1\ 1\ 1\dots\\ 
 & X_3\Rightarrow & 2\ 2\ 1\ 1\ 1\ 1\dots\\
 &  & 2\ 2\ 2\ 1\ 1\ 1\dots\\ 
 &  & 2\ 2\ 2\ 2\ 1\ 1\dots\\ 
 &  &  \dots\\
 &  &  \dots\\ 
 & X_4\Rightarrow & 1\ 1\ 1\ 1\ 1\ 1\dots
\end{matrix}$$
Starting from $X_3$ output may take indefinite time before stabilizing to $Y=1/O_{S1}$

An algorithm to find if disturbance can be decoupled and possible controller for the same is given as follows:\\
Without loss of generality, let the variables present in the output equation be first $s$ variables. Let, $X_{1-s}$ be the set of all states generated from variables $x_1$ to $x_s$ (referred to as output friendly subspace) $X_{1-s}:=\{ \ltimes_{i=1}{s}x_i\}$. Let, $O_{si}$ be set of all states corresponding to output $i$ i.e. $O_{si}:=\{x_{sj}|x_{sj}\in X_{1-s} \ \& \ H(x_{sj})=\delta_{2^p}^i \}$ for $i\in \{1,\dots ,2^p\}$ with slight abuse of notation. Let, $L_s:=$ state-transition matrix corresponding to output friendly variabls only.

\begin{algorithm}
\caption{Disturbance Decoupling in Iteration}\label{alg:Algo_DD_1}

\begin{enumerate}

    \item Divide the states that appear in the output  equations into $2^p$ groups corresponding to each possible output, such that every state from a group has the same output.
    \item Divide state transition matrix $L_s$ into $2^m$ equal blocks, where each block corresponds to one of $2^m$ possible input combinations ($L_{su}^1$ to $L_{su}^{2^m}$ of size $2^s\times 2^{n+d}$).
    \item Divide each $L_{su}^i$ block into $2^s$ sub-blocks, corresponding to all possible combinations of output friendly variables ($L_{suo}^{i_1}$ to $L_{suo}^{i_{2^s}}$ of size $2^s\times 2^{n-s+d}$).
    \item For every index $k$ of $L_{suo}^{i_k}$, list out all the possible next states (entries from $L_{suo}^{i_k}$) as $x_{sk}^{i+}\ \forall i\in \{1,\dots ,2^m\}$ (check for all possible next\_states).
    \item Representing elements of $X_{1-s}$ by the corresponding ordinal Boolean vectors $\delta_{2^s}^k$, define $C_k:=\{set\ of\ possible\ control\ strategies\ for\ DD\}$, then $i\in C_k$ iff $\forall x_{ns}\in x_{sk}^{i+}$, for $x_{ns}\in O_{sl}$ for same $l\in \{1,\dots ,2^p$ i.e. all the possible next\_states $(x_{ns})$ should belong to same output where, $i=\delta_{2^m}^i$
    \item Disturbance decoupling is possible if, $C_k\neq \{\phi \}\ \forall k\in \{1,\dots, 2^s\}$; and the state feedback controller can be constructed as 
    $M_x = \delta_{2^m}[e_1\in C_1 \dots e_{2^{n-s}}\in C_1,\ e_{2^{n-s}+1}\in C_2 \dots e_{2\times 2^{n-s}}\in C_2,\dots \dots, e_{(2^{s-1}+1)\times 2^{n-s}}\in C_{2^s} \dots e_{2^s \times 2^{n-s}}\in C_{2^s}]$ i.e. for every $x_{sj}\in X{1-s}$, for $j\in \{1,\dots,2^s\}$ for every complete state formed by $x_{sj}\ltimes (\ltimes_{i=n-s+1}{x_i})$ $C(x_{sj}\ltimes (\ltimes_{i=n-s+1}{x_i}))\in C_j$ where $C(z)$ is control corresponding to state $z$.
\end{enumerate}

\end{algorithm}

Since the algorithm works on a finite set (argument to algorithm is a finite set) it will terminate with one of two possibilities,
\begin{itemize}
    \item Disturbance can not be decoupled from output.
    \item Disturbance can be decoupled with a well defined control structure.
\end{itemize}

\textbf{Theorem 11:} System given by $$x^+=Lux\xi$$
$$y=Hx$$
can be decoupled from disturbance in output if the algorithm \ref{alg:Algo_DD_1} returns all non-empty sets $C_k\ for\ k\in \{1,\dots,\ 2^s\}$.

\textbf{Proof:} The algorithm can be looked into as a one step look-ahead process, where control is decided such that the output for \textit{next state} remains unaffected by the disturbance and non-output friendly variables. Since this step is performed at every state, the disturbance remains decoupled.\hfill $\square$ 

Note here that the disturbance is decoupled from the output in the sense that the state feedback suggested by the algorithm removes any randomness or stochasticity from $(state,\ input)\mapsto (output)$ map and creats a completely deterministic $(s,i)\mapsto (o)$ map.

This is in agreement with the literature available on disturbance decoupling of $BCN$. In actuality though, what is desired usually is for system to follow certain behaviour in its output. For this controllability of output, the state transition matrix $L$ (system matrix) along with output matrix $H$ need to satisfy some additional conditions as follows:
\begin{itemize}
    \item For any output to remain constant, the corresponding output set $O_s$ must remain invariant.
    
    If every output set is not invariant, then the output can be controlled without variation only for the invariant output sets. Therefore at-least the desired output needs to have the invariant $O_s$.
    \item The desired output\_invariant\_set needs to be reachable from every state.\\
    \textit{Strictly/ Definitely}: Reachable within a predefined number of intermediate states.\\
    \textit{Cleanly}: Directly, without any intermediate states.\\
    \textit{Definitely but not cleanly}: With some intermediate (predefined number of) states with not necessarily equal to the desired output.\\
    \textit{Indefinitely}: Reachable with certain probability (iteration dependent).
\end{itemize}

\textbf{Note:} Effect of disturbance in Boolean systems is much more adverse compared to continuous systems as it can change the course of the system completely. If DD is possible, then unlike continuous systems, its effects can be eliminated completely usually in a small number of iterations.

Existence of elements of a block belonging to a single output\_set is only a necessary condition leading to DD in mapping sense, which generally falls short of required behaviour. To add stronger control over the behaviour along with disturbance decoupling, additional conditions in the form of invariance and reachability are required. One way to achieve this is to migrate from output\_set based behaviour to output\_invariant\_set based behaviour in analysis and control design. If, a digraph is constructed with all possible output\_sets as verices numbered according to Boolean output vectors, i.e. $\mathcal{V}:=\{\delta_{2^p}^1, \dots, \delta_{2^p}^{2^p}\}$ or $\mathcal{V}:=\{1, \dots, 2^p\}$ for simplicity of notation and directed edges $(i,j)\in \mathcal{E}$, connecting vertices if under any control there is a possibility of transition from one $Os$ to another, with a possibility of multiple edges from a single vertex corresponding to a single control.

\begin{figure}[t]
\centering
\includegraphics[scale=0.55]{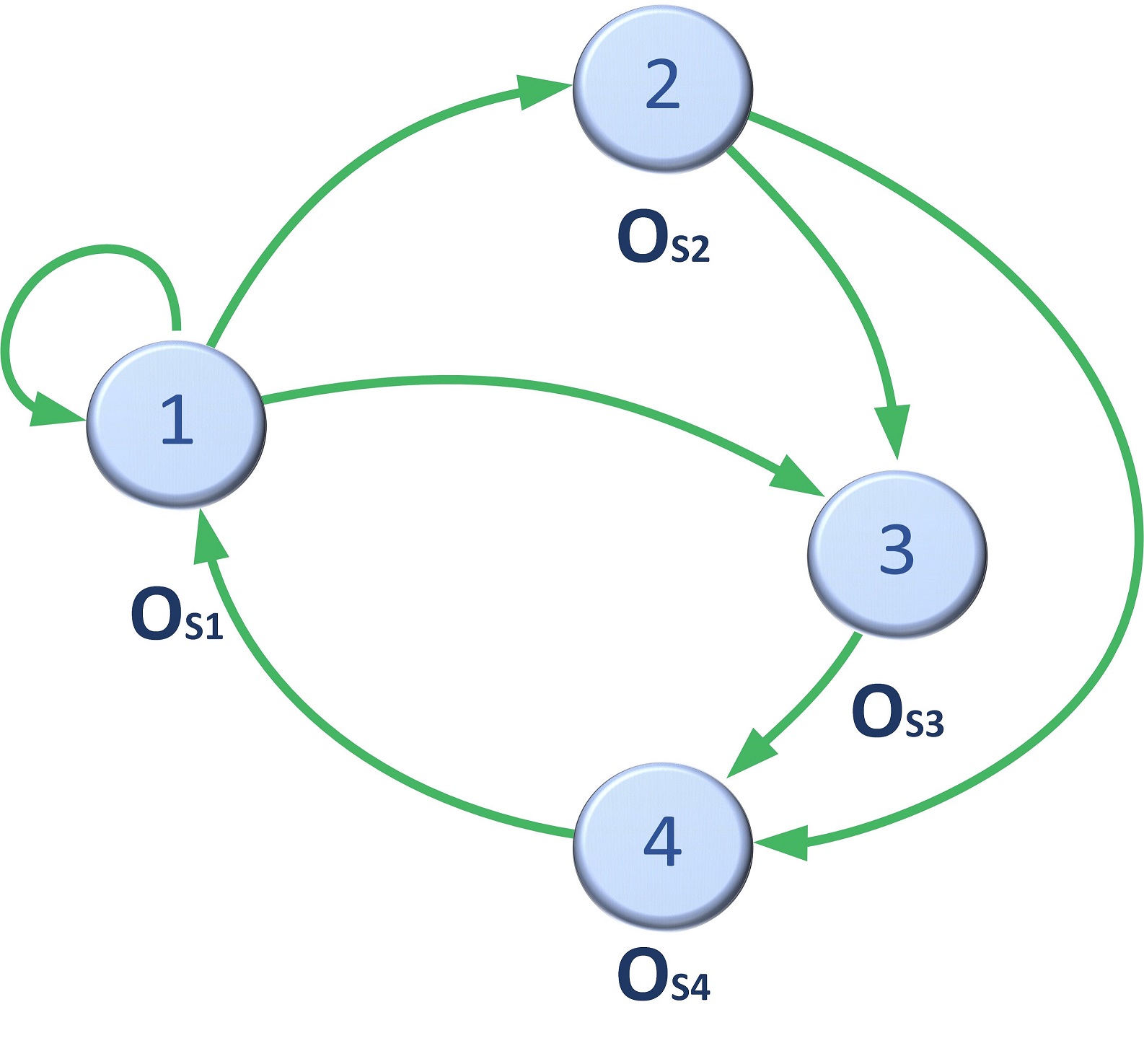}
\caption{Output set digrap}
\label{output_set_reachability}
\end{figure}

For invariance of the output\_set there must be a recurring edge or self-edge to the vertex which is traversed without any uncertainty every time the appropriate control is selected. In other words, for an $O_s$ to be invariant, all the states belonging to it must, under some input, transition to some state in the same $O_s$. Output invariance makes sure that the output remains constant in presence of disturbances. If $O_s$ is invariant then the corresponding output $O_j$ can be made disturbance free, Invariance of $O_{sj}$ is not sufficient though, $O_{sj}$ needs to be reachable. Disturbance can be decoupled, in a stronger sense, if all $O_{sj}$ are invariant and reachable.

Algorithm \ref{alg:Algo_DD_1} can be modified to check for $O_s$ invariance and reachability.

\begin{algorithm}
\caption{Disturbance Decoupling in Iteration Modified}\label{alg:Algo_DD_2}

\begin{enumerate}
    \item Steps (1) to (4): same as Algorithm \ref{alg:Algo_DD_1}
    \setcounter{enumi}{4}
    \item Representing elements of $X_{1-s}$ by the corresponding ordinal Boolean vector $\delta_{2^s}^k$\\
    \hspace{1cm} Repeat for $k\in \{1,\dots,2^s\}$\\
    \ \ Repeat for $i\in \{1,\dots,2^m\}$\\
    $O_{k+}^i=\{l\ |\ \exists x_{ns}\in x_{ns}^{i+}\ such\ that\ x_{ns}\in O_{sl}\}$\\
    $O_{k+}^i$ is the set of outputs of all the possible next\_states of state $k$ with control input $i$. Let, $O_{k+}=\{O_{k+}^i\ |\ i\in \{1,\dots, 2^m\}\ for\ k\in \{1,\dots ,2^s\}$
    \item \textit{State feedback control selection:}\\
    The controller can be constructed as a matrix of $2^s$ blocks, a block of size $(2^m\times 2^{n-s})$, corresponding to every state in $X_{1-s}$. The feedback law is decided as follows:
    \begin{itemize}
        \item \textbf{a. Control for DD in mapping:}\\
        For all available $k\in \{1,\dots, 2^s\}$\\
        $C_k=\{i\ |\ \forall O_{k+}^i\in O_{k+}\}$ such that $O_{k+}^i$ is singleton
        \item \textbf{b. Control for invariant DD:}\\
        For all available $k\in \{1,\dots, 2^s\}$\\
        $C_k=\{i\ |\ \forall O_{k+}^i\in O_{k+}\}$ such that $O_{k+}^i$ is singleton set and $H(k)\in O_{k+}^i$, where $H(k)$ indicates the output at state $k$.
        \item \textbf{c. Control for reachability (clean):}\\
        To reach $O_{sl}$, for all available $k\in \{1,\dots, 2^s\}$\\
        $C_k=\{i\ |\ \forall O_{k+}^i\in O_{k+}\}$ such that $O_{k+}^i$ is singleton set and $l\in O_{k+}^i$
        \item \textbf{d. Control for reachability (definite):}\\
        To reach $O_{sl}$, provided that at-least one directed path exists,\\
        if $k$ is current state and $O_{sn}$ is the expected next output\_set vertex in the directed graph, then\\
        $C_k=\{i\ |\ \forall O_{k+}^i\in O_{k+}\}$ such that $O_{k+}^i$ is singleton set and $n\in O_{k+}^i$
        \item \textbf{e. Control for reachability (indefinite):}\\
        To reach $O_{sl}$, provided that at-least one directed path exists,\\
        if $k$ is current state and $O_{sn}$ is the expected next output\_set vertex in the directed graph, then\\
        $C_k=\{i\ |\ \forall O_{k+}^i\in O_{k+}\}$ such that $n\in O_{k+}^i$
    \end{itemize}
\end{enumerate}

\end{algorithm}

Note here that $C_k$ is the set of all the possible control actions when in state $k$. $i\in C_k$ implies $i\in \delta_{2^m}$ or $i:=\delta_{2^m}^i$. If for any $k$, the returned $C_k=\{\phi \}$  then this indicates that no control action is possible to achieve desired behaviour. Notice also, that the control selection for case (e) is online process, i.e. in general the control signal can not be determined/planned in advance. Therefore knowledge of current state becomes necessary and observer may be utilized.

\textbf{Theorem 12:} For $BCN$ in (9), (a) DD in mapping (b) Invariant DD (c) Clean Reachability (d) Definite Reachability (e) Indefinite Reachability, is achieved if \textit{Algorithm 2} returns all $C_k=\{\phi \}$ for respective requirements.

\textbf{Proof:}(a) Same as proof \textit{Theorem 11}\\
(b) Similar to (a), one-step look-ahead is performed with extra condition that the next\_state output is same as the current output.\\
(c),(d) \& (e) follow from prepositions 1, 2 \& 3 respectively.\hfill  $\square$

Looking at the notions of the reachability presented, it can be observed that the methods available (in the literature) require the states of the output sub-system to follow clean reachability i.e. every state of the output sub-system needs to cleanly reach at least one state. This requirement is quite conservative as it leaves out the possibility of indefinite reachability entirely, even though it is much more inclusive and analogous to reachability in the continuous domain. To include indefinite reachability in DD, the following invariant set construction is utilized:

\begin{algorithm}
\caption{Disturbance Decoupling: Invariant Set}\label{alg:Algo_DD_Inv}

\begin{enumerate}
    \item In the output sub-system construct the largest controlled invariant set under the system dynamics with clean reachability, i.e.\\
    $S_1=\{x_s\in X_{1-s}\ |\ \exists i\in \{1,\dots,2^m\}\ \&\ L'u_ix_s\in S_1\}$
    \item Construct the sub-system states not included in set in step 1, set of all the states that cleanly reach the set in 1, i.e.\\
    $S_2=\{x_s\in X_{1-s}/S_1\ |\ \exists i\in \{1,\dots,2^m\}\ \&\ L'u_ix_s\in S_1\}$
    \item Construct from states not included in sets in either steps 1 or 2, the set of all the states that cleanly reach $[set\ in\ 1\  \bigcup \ set\ in\ 2]$, in other words, two sets defined in 1 or 2, i.e.\\
    $S_3=\{x_s\in X_{1-s}/(S_1\bigcup S_2)|\exists i\in \{1,\dots,2^m\}\ \&\ L'u_ix_s\in S_1\bigcup S_2\}$
    \item \dots
    \item \dots
\end{enumerate}
Continue till remaining states can not be classified.

\end{algorithm}

If the $set\_of\_remaining\_states:=S_{rs}=\{\phi\}$ (i.e. set containing all the states that can not be classified is an empty set), then the disturbance can be decoupled.

Note that, from step 2 onwards the notion of clean reachability is more inclusive, as it is used in the context of a set and is equivalent to indefinite reachability with restriction.\

\begin{figure}[t]
\centering
\includegraphics[scale=0.25]{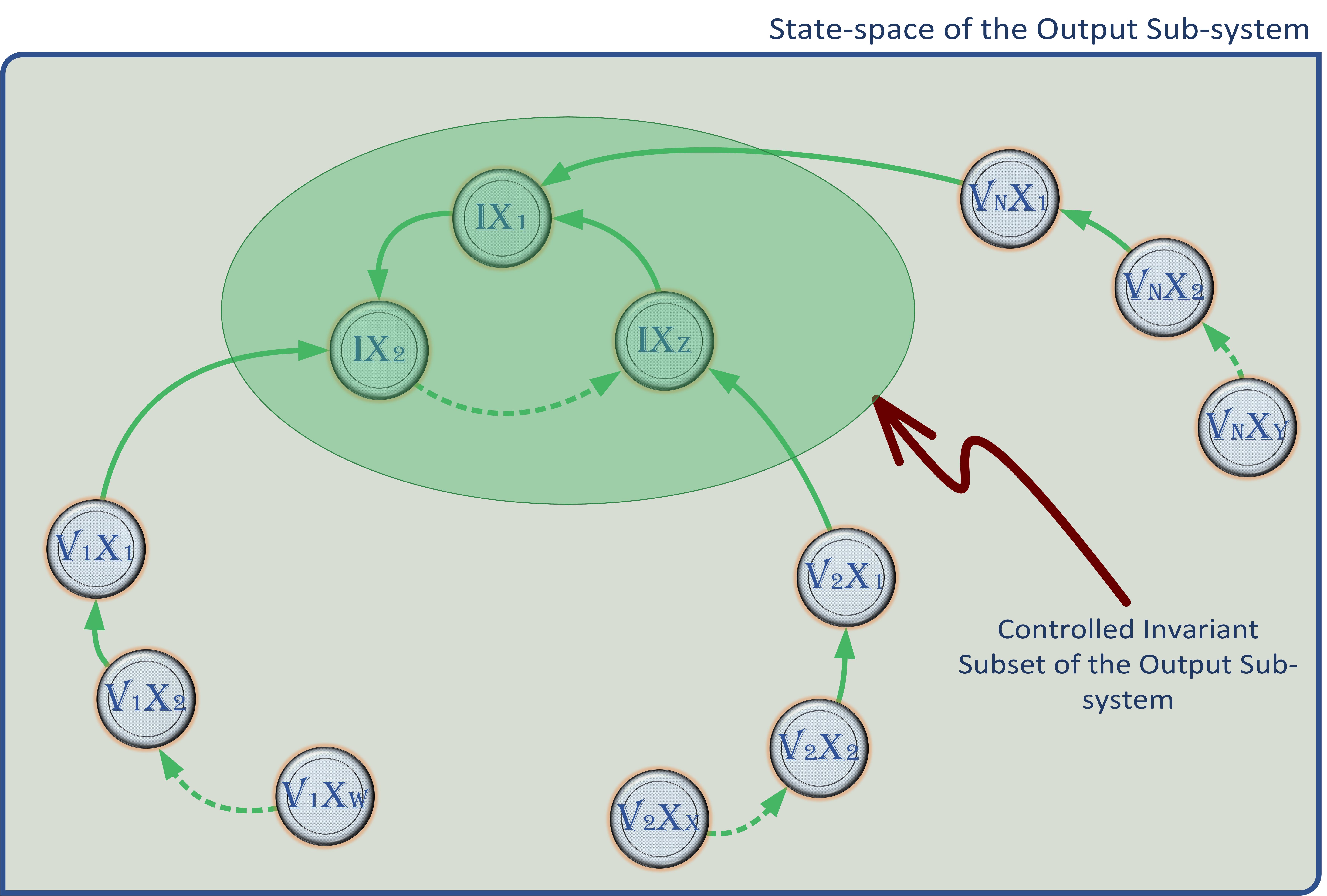}
\caption{Control invariant subset}
\label{control_invariant_subset}
\end{figure}

If no states of the sub-system remain unclassified then the disturbance can be decoupled in at-most $S_D$ steps, where $S_D$ is maximum of the number of evolutions required by all the states to reach invariant set. Based on the control law under which the invariant behaviour is achieved, the controller can be classified as pinning controller or state-feedback controller.

Simply put this algorithm constructs definitely reachable sets that reach the controlled invariant sub-set in number of evolutions, upper bonded by some integer. If the algorithm terminates in its first step, i.e. after construction of $S_1$, with $S_{rs}=\{\phi\}$, the system satisfies the necessary and sufficient condition. It therefore can be decoupled from the disturbance in the sense of literature available. Otherwise, if it terminates at any higher step with $S_{sr}=\{\phi\}$, the disturbance can not be decoupled as per the literature, but it is decoupled in the mapping, with the system trajectory to the controlled invariant sub-set of the output sub-system.

The algorithm \ref{alg:Algo_DD_Inv} can be modified as Algorithm \ref{alg:DD_Fin_Iter_2} to work on the state transition matrix $L'$ of the output sub-system, instead of the set of the output sub-system states.\footnote{The state transition matrix of the output sub-system i.e. $L'$ can be built from the state transition matrix of the entire system i.e. $L$, uniquely}

\textbf{Theorem 13:} If algorithm \ref{alg:Algo_DD_Inv} classifies all the states into sets with $S_{rs}=\{\phi\}$, then the disturbance can be decoupled in mapping for controlled invariant output subset.

\textbf{Proof:} If the algorithm terminates with $S_{sr}=\{\phi\}$, this implies that all the states of the system (output sub-system) are classified into some set $S_i$. A reachability graph can be constructed for clean reachability as follows:\\
$\mathcal{V}:=\{S_i|i\in \{1,\dots ,k\}\}$, where $k$ is the number of sets created by the algorithm except $S_{rs}$ and an edge set $\mathcal{E}$ with $(S_i,S_j)\in \mathcal{E}$ if $\forall x_{S_i}\in S_i,\ \exists u_l\in \delta_{2^m}$ such that $L'u_lx_{S_i}\in S_j$\\
As per construction of the sets,\\
$\forall x_{S_{i+1}}\in S_{i+1},\ \exists u_l\in \delta_{2^m}$ such that $L'u_lx_{S_{i+1}}\in S_i$\\
$\therefore (S_{i+1},S_i)\in \mathcal{E}\ \forall\ i\in \{1,\dots ,k-1\}$\\
$\therefore \mathcal{E}:=\{(S_{i+1},S_i)\ |\ i\in \{1,\dots,k-1\}\}$

From the digraph it is obvious that there is directed path from every set $S_i$ to $S_1$, hence $S_1$ is definitely reachable from every set. Since, $S_1$ can be decoupled from disturbance, system can be disturbance decoupled in mapping restricted to $S_1$.\hfill $\square$

The disturbance is decoupled in at-most $S_D$ steps, because no matter the initial state, after $S_D$ evolutions the system will be in the controlled invariant sub-space; which is not affected by the disturbances.\footnote{If a controller can be designed as a feedback (state/ output)/ pinning control then it is said that the DDP is solvable using feedback (state/ output)/ pinning control}

\textbf{Controlled Invariant Sub-set:} A sub-set is controlled invariant if under the action of some control law, the system trajectory remains in the sub-set ones entered. In terms of state transition matrix, this means that the corresponding columns of the state-transition matrix are the elements of the same sub-set.

In case when there is no possibility of disturbance decoupling, the best that can be done is to provide an estimate or probability that the disturbance will be decoupled after any number of evolutions. As the number of evolutions increase, the possibility that the disturbance is decoupled increases and approaches 1 as the the number of iterations approach $\infty$ (provided there exists a universally indefinitely reachable invariant sub-set of output sub-system).

An interesting observation can be made on algorithm \ref{alg:DD_Fin_Iter_2}\\
Let $x_i:=x(i)$, $\xi_i:=\xi (i)$ and $\tilde{L}:=LM_x\psi_n$, then
$$x(1)=\tilde{L}x_0\xi_0$$
$$x(2)=\tilde{L}x_1\xi_0=\tilde{L}(\tilde{L}x_0\xi_0)\xi_1=(\tilde{L})^2x_0\xi_0\xi_1$$
in general,
$$x(k)=(\tilde{L})^kx_0\tilde{\xi}_k$$
where, $\tilde{\xi}_k:=\xi_0 \cdot \xi_1 \dots \xi_{k-1}$\\
Suppose, the algorithm \ref{alg:DD_Fin_Iter_2} returns $S_{rs}=\{\phi\}$ and $k$ is the largest value of the set index generated by the algorithm, i.e. $k=max\{i\,|\,S_i\neq \{\phi\}\}$. In other words, the disturbance is decoupled in at-most $k$ evolutions ($k$ can be considered as the largest distance from the invariant set).\\
For the system in which all the state variables contribute to the output $\tilde{L}_k:=(\tilde{L})^k$ has the following structure,\\
let $S_1:=\{Controlled\ invariant\ set\}$, divide $\tilde{L}_k$ into  $2^n$ blocks as $\tilde{L}_k=[\tilde{L}_k^1\ \tilde{L}_k^2 \dots \tilde{L}_k^{2^n}]$, then the blocks corresponding to the invariant set $S_1$ are of rank 1, i.e. for $i\in \{S_1\}$ $rank[\tilde{L}_k^i]=1$. The columns of the remaining blocks are the elements of the invariant set, i.e. for $i\notin \{S_1\}$  $Col(\tilde{L})_k^i\in \{S_1\}$, where $Col(\cdot)$ indicates the column set of a matrix.

For the case where not all the variables appear in the output equation, the same observations apply to $\tilde{L}_k^o$, where $\tilde{L}_k^o:=(\tilde{L}^o)^k$ and $\tilde{L}^o$ is the state transition matrix of the output sub-system built from $\tilde{L}$. This observation provides an interesting insight on how the \textit{DD in iteration} works.

\textbf{E.g. 7.} Let $x^+=Lu_1x_1x_2\xi_1$ and $y=h(x_1,x_2)$ where $h(\cdot)$ is any Boolean function. Let, 
$$L=\delta_4[1\ 2\ 3\ 4\ 1\ 2\ 3\ 3\ 3\ 4\ 1\ 3\ 4\ 4\ 2\ 3]$$
then the algorithm \ref{alg:DD_Fin_Iter_2} will return, 
$$S_1=\{3,4\},\ S_2=\{1,2\}$$
$$C_1=\{2\},\ C_2=\{1\},\ C_3=\{2\},\ C_4=\{1\}$$
$$\therefore M_x=\delta_2[2\ 1\ 2\ 1]$$
$$\tilde{L}_1=LM_x\psi_2=\delta_4 [3\ 4\ 3\ 4\ 4\ 4\ 3\ 3]$$
For $i\in \{S_1\}=\{3,4\}$, $rank[\tilde{L}_k^i]=rank[\tilde{L}_1^i]=1$. $k=1$ because starting from any state, $\{S_1\}$ can be reached in at-most 1 evolution. For $i\notin \{S_1\} \rightarrow i\in \{1,2\}$, $Col(\tilde{L}_1^i)\in \{S_1\}$.

\textbf{E.g. 8.} Let,
$$L=\delta_4[3\ 3\ 3\ 4\ 1\ 2\ 1\ 2\ 3\ 4\ 3\ 3\ 3\ 3\ 2\ 4]$$
therefore,
$$S_1=\{3\},\ S_2=\{1,2\},\ S_3=\{4\}$$
$$C_1=\{1\},\ C_2=\{2\},\ C_3=\{2\},\ C_4=\{1\}$$
$$M_x=\delta_2[1\ 2\ 2\ 1]$$
$$\tilde{L}=LM_x\psi_2=\delta_4 [3\ 3\ 3\ 3\ 3\ 3\ 1\ 2]$$
$$\tilde{L}_2=(\tilde{L})^2=\delta_4 [3\ 3\ 3\ 3\ 3\ 3\ 3\ 3\ 3\ 3\ 3\ 3\ 3\ 3\ 3\ 3]$$
Hence, for $i\in \{S_1\}=\{3\},\ rank[\tilde{L}_k^i]=rank[\tilde{L}_2^i]=1$. For $i\notin \{S_1\}\rightarrow i\in \{1,2,4\},\ Col(\tilde{L}_2^i)\in \{S_1\}$
\subsection{Disturbance Decoupling Utilizing Output Equation}
The results provided so far address various possibilities by expanding the notion of DD without utilizing actual structure of the output equation, as is the case with much of the literature. This restriction may be essential or may be unnecessary depending on the requirements. E.g. to achieve $BN-BCN$ equivalence in state transition in presence of disturbances the output equation can not be utilized, on the other hand for $BN-BCN$ equivalence in output sequence in presence of disturbances the output equation plays an important role. In \cite{zou2017state} authors have tried to utilize the actual output equation or the $H$ matrix, but the results obtained are inconsistent. E.g. solving for the example (using STP toolbox) considered in the work with two different disturbance sequences yields two different outputs, namely\\
for $\xi_1: 1\ 1\ 0\ 0\ 0\ 1\ 1\ 1\ 0\ 1$ (same as the first 10 elements of $\xi_1(t)$ from \cite{zou2017state}); the resulting output sequence turns out to be $O_1: 1\ 1\ 1\ 0\ 0\ 0\ 1\ 0\ 1\ 0$\\
but for $\xi_2: 1\ 0\ 0\ 0\ 0\ 0\ 1\ 0\ 0\ 0$; the resulting output sequence turns out to be $O_2: 1\ 1\ 0\ 0\ 0\ 0\ 0\ 1\ 0\ 0$\\ which is a mismatch compared to what is claimed.

In this section results on DD utilizing output equations are presented along with results on variable reference tracking feedback construction analogous to classical control.

The previously presented the notions of disturbance decoupling are applicable in this case too, with a slight modification, in the definition of the map
$$H_L:(state,input)\mapsto (next\_output)$$
One thing obvious from this definition is $(DD\ invariant\ in\ y)\rightarrow (DD\ in\ mapping)\rightarrow (DD\ in\ iteration)$, which will be helpful in implementing different structures of controls to steer the system behaviour.
\begin{eqnarray*}
y^+ = HLux\xi_d 
\end{eqnarray*}
State feedback:
$$y^+=M^Ox\xi_d\ where\ HLM_x\psi_n=M^O$$
For disturbance to be decoupled $M^O$ should have following structure:
\begin{itemize}
    \item Divide $L$ into $2^n$ blocks of $2^d$ columns each
    \item Then each block should have rank 1
\end{itemize}
Let $\tilde{L}=LM_x\psi_n$, this matrix has structure of the form
\begin{multline}
    \tilde{L}=[Blk\ (2^{i_1}+1)\ of\ L\ Blk\ (2^{i_2}+2)\ of\ L\ \dots\ \\
    Blk\ (2^{i_{2^n}}+2^n)\ of\ L]
\end{multline}
where $i_k\in \{0,\dots,2^m-1\}$. Dividing $M^O$ into $2^n$ blocks of size $2^p\times2^d$, each block should be of rank 1.\\
We have,
$$M_{ij}=Row_i(H)\times Col_j(\tilde{L})$$
For $M^O$ to satisfy the rank condition (more specifically, blocks of $M^O$ to satisfy rank condition), $2^d$ consecutive columns starting from $2^k+1$ should have 1 at $k^{th}$ row.\\
Let $Col(H)^r:=\{j|Col_j(H)=y_r\}$, where $y_r-r^{th}$ output value\\
then $Col(H)^r$ indicates set of column indices in $r^{th}$ row of $H$ where matrix entry is 1. Therefore,
$$ M^O_{ij}=1$$
$$\text{iff}\ [Col_j(\tilde{L})]_m=1\ \text{and}\ m\in Col(H)^r$$
This means, depending upon the structure of $H$, blocks of $M^O$ will satisfy the rank condition iff,
$$\text{for}\ Col_j(\tilde{L}),\ (2^k+1)\leq j\leq (2^k+2^d),\ 0\leq k\leq (2^n-1)$$
$$e_O[Col_j(\tilde{L})]\in Col(H)^{i_k}$$
$$\text{where}\ e_O(\mathcal{V}_B):=\{l\ |\ \mathcal{V}_{B_l}=1\}$$
$$\text{where,}\ \mathcal{V}_B-\ \text{any Logical vector}$$
Structure of $H$ imposes conditions on the structure of $\tilde{L}$ that,
\begin{multline*}
    e_O[Col_j(\tilde{L})]\in Col(H)^r\ \forall j,\\ (2^k+1)\leq j\leq (2^k+2^d),\ 0\leq k\leq (2^n-1)
\end{multline*}
i.e. for every block of size $(2^p\times 2^d)$ of $\tilde{L}$
$$Col(Blk_i\tilde{L})\subseteq \{\delta_{2^n}^l|l\in Col(H)^r\}$$
Since, $\tilde{L}$ has a structure as indicated by (1), this means a particular structure of $\tilde{L}$ will impose restrictions on $L$.\\
$\tilde{L}$ matrix depends upon $M_x$ and $L$ matrices, to follow conditions of $\tilde{L}$, $L$ needs to satisfy following condition:
\begin{itemize}
    \item Divide $L$ matrix into $2^m$ blocks of $2^(n+d)$ columns each. Assign indices $(Blk_iL)\ ,\ 1\leq i\leq 2^m$.
    \item Divide each $(Blk_iL)$ into $2^n$ sub-blocks of $2^d$ columns each. Assign indices $(Blk_iL)^j\ ,\ 1\leq j\leq 2^n$
    \item For every sub-block index $j$, there should be at-least one block index $i$ such that
    $$Col[(Blk_iL)^j]\subseteq \{\delta_{2^n}^l|l\in Col(H)^r\}$$
    for some $y_r$
\end{itemize}

It can be further illustrated as follows:\\
Generally, disturbance decoupling means elimination of disturbance from output dynamics, for a $BCN $ defined by
\begin{eqnarray}
    x^+ &=& Lux\xi\\
    y &=& Hx
\end{eqnarray}
following expression holds
$$y^+=Hx^+$$
$$y^+=HLux\xi$$
if input is chosen as state feedback defined by $M_x$ then,
$$y^+=HLM_x \psi_n x\xi$$
Let $\tilde{L}:=LM_x\psi_n$ then,

\begin{figure}[!h]
\centering
\includegraphics[scale=0.42]{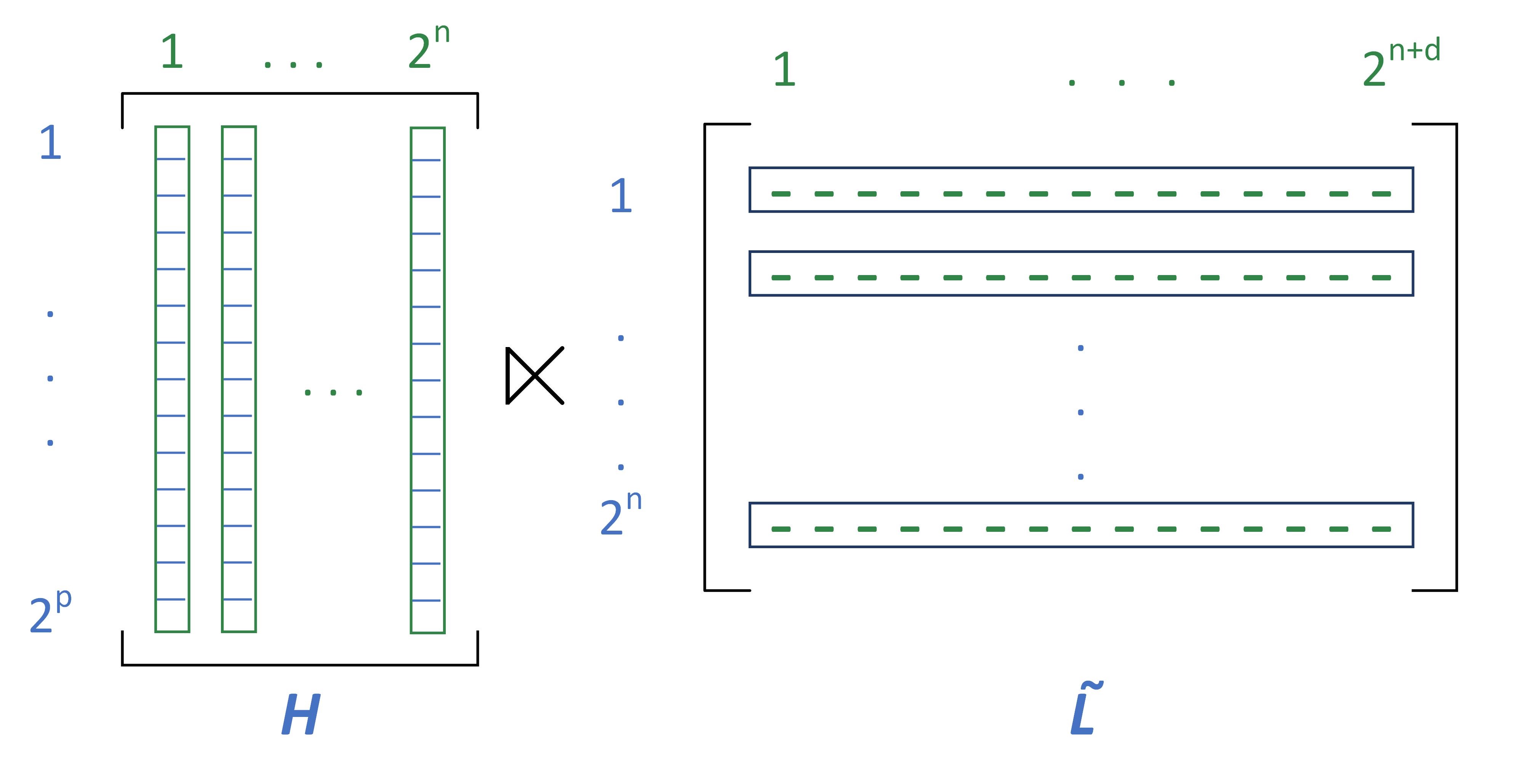}
\label{output_matrix_multiplication}
\end{figure}

This multiplication can also be interpreted as $$H\tilde{L}=\sum_{i=1}^{2^n}Col_i(H)Row_i(\tilde{L})=\sum_{i=1}^{2^n}M_i$$ where $M_i$ represent the rank 1 matrices. If $Col_i(H)$ is denoted by $H_i$ and $Row_i(\tilde{L})$ by $\tilde{L}^i$, then $$H_i\tilde{L}^i=\begin{matrix}
1\\ 
\\ 
j\\ 
\\ 
2^p
\end{matrix}
\begin{bmatrix}
0 & 0 & \dots & 0\\ 
\dots &  &  &  \\ 
[ &  &  & ] \\ 
\dots &  &  &  \\ 
0 & 0 & \dots & 0  
\end{bmatrix}$$
$i^{th}$ row of $\tilde{L}$ appears as the $j^{th}$ row of $H_i\tilde{L}_i$ where $j$ is the position (index) of element $1$ in the $i_{th}$ column of $H$, $H\tilde{L}=\sum H_i\tilde{L}_i$ of dimension $(2^p\times 2^{n+d})$. There are $2^p$ possible outputs, consider columns of $H$, that correspond to the same output. Let, $H^r:=\{ j|Col_j(H)=y_r\}$, be the set of columns for which output is $y_r$ (more precisely, $1$ is at $r^{th}$ index). The $i^{th}$ row of $H\tilde{L}$ is formed by summing over all $\tilde{L}_k$ such that $k\in H^i$. Since no two $\tilde{L}$ can be identical, the following condition must be satisfied:\\
Define $\tilde{L}_{ir}:=\sum {}{k}\tilde{L}_k,\ k\in H^i$\\
Divide $\tilde{L}_{ir}(1\times 2^{n+d})$ into $2^n$ blocks each with $2^d$ elements, then for disturbance to not affect the output $\rightarrow$ all the elements in any block should be identical i.e. either $0$ or $1$. \\
Furthermore, system can be divided into state variables that appear in the output expression (output friendly) and state variables that do not, as well as disturbance.

Let, $S$:- the number of output friendly variables, $(n-s):-$ number of not output friendly variables. Then, divide $\tilde{L}_{ir}(1\times 2^{s+(n-s)+d})$ into $2^s$ blocks, each with $2^{n-s+d}$ elements, then for disturbance to not affect the output $\rightarrow$ all the elements should be identical. (i.e. every element of a sub-block should belong to the same output group for some input). One obvious advantage of this method over \cite{yang2013controller} is that it utilizes the output equation of the system. The difference is evident from the necessary condition, the method presented here, requires $\tilde{L}_{ir}$ to satisfy the given condition, but the individual rows in the sum need not. Contrary to this, in \cite{yang2013controller} every row of $\tilde{L}$ individually needs to satisfy the presented condition.

\textbf{E.g. 9.} Suppose $H=\delta_2[1\ 2\ 1\ 2]$ i.e. $H:f_y(x_1,x_2)$ and $\tilde{L}_s=\delta_4[1\ 3\ 1\ 1\ 2\ 4\ 4\ 2\ 3\ 1\ 3\ 3\ 2\ 2\ 2\ 4]$ i.e.  $\tilde{L}_s:f_x(x_1,x_2,x_3,x_4)$. The methods existing in literature will classify the given system as unsuitable for disturbance decoupling; however it is clearly DD solvable using the method presented here.\footnote{Work presented here is generalization of the existing ideas for more inclusivity. Its utility is application dependent. Stricter demands may require a more conservative approach.}
\subsection{Disturbance Decoupling with Output Feedback}
Let system dynamics be given by
$$x^+=Lux\xi$$
with output feedback $M_y$ the control input will take the form $u=M_yy=M_yHx$ resulting in
$$x^+=LM_yH\psi_nx\xi=\tilde{L}x\xi$$
for this system to reject the disturbances the following condition must hold:
\begin{itemize}
    \item Divide $\tilde{L}$ matrix into $2^n$ blocks, then every block $Blk_i(\tilde{L})$ has rank 1.
\end{itemize}
The matrix $\tilde{L}$ can be represented as,
\begin{multline}
    \tilde{L}=[Blk_{Col_1(\tilde{H}_F)}(L)\ |\ Blk_{Col_2(\tilde{H}_F)}(L)\ |\ \dots\ \\ 
    \dots Blk_{Col_{2^n}(\tilde{H}_F)}(L)]
\end{multline}
In other words, $2^n$ blocks of $L$ are selected based on the output feedback. For disturbance decoupling all such selected blocks need to satisfy the rank condition.

For disturbance decoupling in output i.e. using output equation,
\begin{eqnarray}
    y^+=H\tilde{L}x\xi=\tilde{H}_Lx\xi
\end{eqnarray}
divide $\tilde{H}_L$ into $2^n$ blocks of size $(2^p\times 2^d)$ each, then each block should have rank 1. This is equivalent to following condition:
\begin{itemize}
    \item Divide $\tilde{L}$ into $2^n$ blocks of size $(2^n\times 2^d)$, then columns of each block belong to the same output group
\end{itemize}
i.e. if
\begin{multline}
    \tilde{L}=[Blk_1(\tilde{L})\ |\ Blk_2(\tilde{L})\  \dots\  Blk_{2^n}(\tilde{L})\ ]
\end{multline}
then, $Col(Blk_i[\tilde{L}])\subseteq O_{sj}$ for some $j\in \{1,\dots,2^p\}$, where $O_{sj}:=\{x\in \delta_{2^n} | Hx=\delta_{2^p}^i$\\
A procedure for finding the output feedback resulting in disturbance decoupling is presented in the following:
\begin{enumerate}
    \item Find all possible state feedback controllers.
    \item Out of these, list the ones for which, all the states belonging to one output group have same control action.
    \item If no such controllers, then DD with output feedback is not possible. Else, the output feedback is given by $M_y$, with $Col(M_y)\subseteq \delta_{2^m}$ and $Col_j(M_y)=u_k$ such that $u_k$ is the common control input corresponding to output\_set $O_{sj}$
\end{enumerate}

\textbf{Remarks:}
\begin{itemize}
    \item Output feedback is difficult to construct and feasible design is less likely to exist compared to the state feedback. Because in output feedback, the control action is common for the entire output\_set, which is a stringent requirement and likely to result in contradiction. Such is not the case with the state feedback.
    \item When number of states and outputs differs largely, especially with small number of outputs, the output feedback can deal with only a slightly disturbed system (i.e. the disturbance affects only a small number of states)
    \item More is the number of outputs, better is the performance of the output feedback in the sense that it can work with worse systems.
\end{itemize}
\textbf{E.g. 10.}
$$\begin{matrix}
L= & \delta_8 [\ 2\ 2\ 4\ 4\ 6\ 6\ 8\ 8\ 1\ 4\ 3\ 5\ 4\ 2\ 3\ 3\\ 
    & \; 1\ 1\ 3\ 4\ 5\ 5\ 8\ 8\ 3\ 3\ 4\ 4\ 5\ 5\ 7\ 7\ ]
\end{matrix}$$
$$H_1=\delta_2[1\ 1\ 1\ 1\ 2\ 2\ 2\ 2]$$
then with output feedback defined by $M_{y1}=\delta_2[1\ 2]$ the resulting $\tilde{L}_1$ that decouples the disturbance is 
$$\tilde{L}_1 = \delta_8 [2\ 2\ 4\ 4\ 6\ 6\ 8\ 8\ 3\ 3\ 4\ 4\ 5\ 5\ 7\ 7]$$
Similarly, for 
$$H_2=\delta_4 [1\ 1\ 2\ 2\ 3\ 3\ 4\ 4]$$
with output feedback defined by $M_{y2}=\delta_2[1\ 1\ 2\ 2]$ the resulting $\tilde{L}_2$ that decouples the disturbance is 
$$\tilde{L}_2 = \delta_8 [2\ 2\ 4\ 4\ 6\ 6\ 8\ 8\ 3\ 3\ 4\ 4\ 5\ 5\ 7\ 7]$$
resulting in same system dynamics.

\section{Strategies for Recunstructability and Observer Design for BCN with Conventional Input Structure}

In literature a couple of notions on observability of $BCN$ are available, due to D. Cheng, et. el \cite{cheng2010analysis} and E. Fornasini, et.el \cite{fornasini2012stateobservers}. The former requires existence of only one control input sequence under which the systems is observable. Whereas in the late case systems needs to observable under all possible inputs for it to be classified as observable (here the term `observable' is used in the sense that for the input sequence in question by observing the output sequence for a finite time the initial state can be uniquely constructed). Referring to these possibilities of observability as observability in weak sense and observabilty in strong sense respectively, one can notice the limitations these approaches hold. Given any application, observability in its weakest sense might be unutilizable on the other hand the observability in its strongest sense might be too restrictive. Therefore it is only well advised to opt for a compromise between the two approaches. Observabilty enjoys a special place in control design as an essential part of the observer design. Still, as shown in \cite{fornasini2012stateobservers}, observability is only a sufficient condition for observer design in BCN and reconstructibility suffices even if the system is not observable, to construct an observer. Though the reconstructability presented in \cite{fornasini2012stateobservers} is a weaker condition than the observability in strong sense, is still might be too conservative for a large number of systems.

This has implications in $BCN$. Any $BCN$ with control signal provided via a conventional method (i.e. state feedback, output feedback or pinning control) transformers to a $BN$. For every $BCN$ that follows the strict necessary and sufficient conditions, there might be a large number of structurally similar networks that fail the test. This provides an incentive to make results available in the literature more inclusive.\footnote{The performance of the included systems may of-course be slack compared to the system that satisfy the stricter conditions, but still may be tolerable.}
 
\subsection*{Conventional Control:}(Time invariant/ non-adaptive control)
Arguably the most important part of any control system is the control law. In open loop systems, the control or the controller acts as an operator that transforms the reference signals to actuator signals. In feedback control the controller reacts to the changing state or output, through control input as state or output feedback. Usually the control law connecting state or output to the input, remains invariant throughout the operation. This static nature of the control law is what is referred to here as the conventional control law that is characterised by a constant $state \mapsto input$ map. Therefore, be it the case of output feedback or state feedback the $state-input$ pair remains constant, i.e. for any $(x_i,u_i)$ and $(x_j,u_j)$, $(x_i=x_j)\Rightarrow (u_i=u_j)$.
 
\subsection*{Strategies for Reconstructibility:}
The reconstructibility of $BCN$ requires \cite{fornasini2015fault} that all the possible $state-input$ cycles of same length produce pairwise different output tuples. Since this requires to be followed for any arbitrary input sequence starting from any initial state, it turns out to be an excessively strict requirement. This follows from the fact that most of the applications utilize conventional control to achieve necessary results. This use of the conventional control induces a nice static $state\rightarrow input$ map. Limiting the possibilities of the input sequence from arbitrary to cyclic with fixed length; therefore the search for the output tuple can be restricted to cyclic $state\rightarrow input$ trajectories of fixed length. This approach too is helpful only in case of a small number of inputs and states. E.g. For a system with $n=4,\ m=1$ the number of possible state feedback is $\underset{16\ times}{\underbrace{2\times 2\times \dots \times 2}} = 2^{16}=65536$, for $n=5,\ m=2$ the number reaches to $\underset{32\ times}{\underbrace{8\times 8\times \dots \times 8}} \simeq 8\times 10^{28}$. Therefore not a feasible approach to check for reconstructibility. This means the search needs to be much more restricted.
 
One possible way to reduce this number is to only look for feedback laws, that yield the required behaviour. Once all such feedback laws are identified, then the reconstructibilty of the $BN$ that results from the corresponding feedback law can be checked.

\subsection*{Reconstructible Output Feedback:}

In absence of knowledge of state, the state feedback becomes meaningless. Searching for reconstructibility in most of the cases is infeasible, in such cases, output feedback may be used to speed up the convergence of the observer. The reasoning behind this approach is that, if following a particular input sequence, the state is somehow constructed then this knowledge can be utilized for state feedback to achieve a greater degree of freedom in controlling the system. The same idea is expressed/formalized in terms of the following algorithm:
 
\begin{figure}[t]
\centering
\includegraphics[scale=0.8]{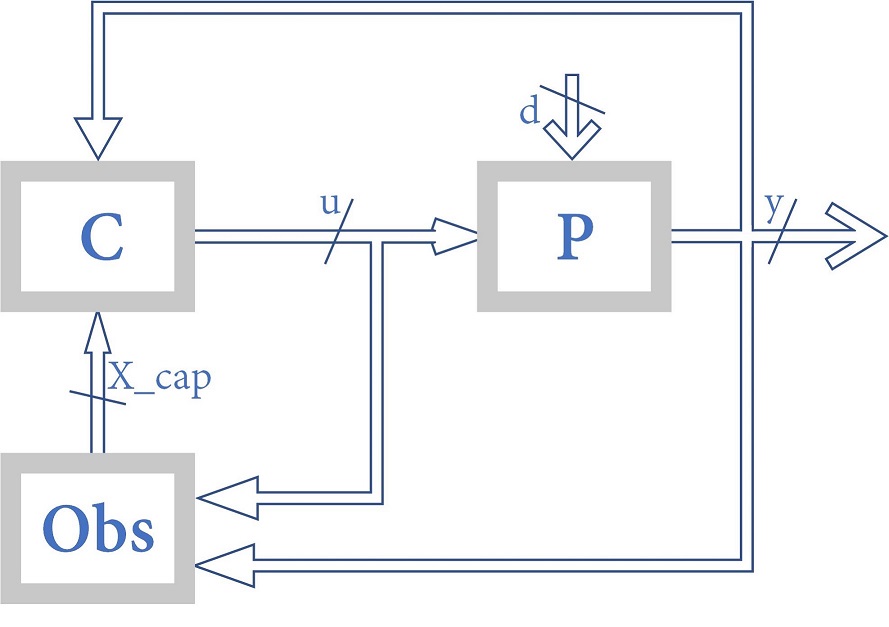}
\caption{Reconstructibility with output feedback}
\label{Reconstructible_output_feedback}
\end{figure}

\begin{algorithm}
\caption{State Reconstructible Output Feedback}\label{alg:St_Recns_OP_FB} 
 
 \begin{enumerate}
     \item Construct from output $y\rightarrow \{possible\ states\}$ for $y_1$ to $y_{2^p}$
     \item For each $y_i$ select an input such that $y_i\rightarrow X_{y_i}^+$ has least number of possible states
     \item Observe the output and apply the corresponding input 
     \item Observe the next output $y_k^{+}$; the state is in 
     $X_E := \{ X_{y_j}^{+} \cap y_k \}$.
     
     If $\left | X_E \right | = 1$, the state is reconstructed. 
     
     If $X_E = \{\phi \}$, fault occurred. 
     
     If $\left | X_E \right | > 1$, Select input according to $y_k$
     \item Go to step (4).
 \end{enumerate}

\end{algorithm}

\subsection*{Possibility of Reconstructibility with Output Feedback:}

As shown earlier, checking all the possible state feedbacks could be a hopeless task, let alone all the possible arbitrary input sequences. Giving rise to following issues:
\begin{itemize}
    \item For arbitrary input the number of possibilities increases faster than exponent
    \item Even with conventional input strategies the number of possibilities may still be large
    \item Representation in matrix form makes search for cycles computationally more expensive
    \item The recursive method proposed in \cite{zhang2016observer} the matrices grow rapidly, making it too unfeasible.
\end{itemize}
These issues could be addressed as follows:
\begin{itemize}
    \item Representing the entire $BCN$ as a digraph with $2^n$ nodes and $2^{n+m}$ edges covering all possibilities of conventional feedback (There might be some repeated edges).
    \item Search for all possible cycles, an algorithm (defth first search ect.) with complexity $O(\left | \mathcal{V} \right | + \left | \mathcal{E} \right |)$ can be used, in this case $O(2^n + 2^{n+m})$, to identify all the cycles much efficiently.
    \item Divide the cycles into reconstructible and non-reconstructible categories.
    \item Of reconstructible cycles, separate out the ones that can be generated through output feedback, by looking for state feedback strategies that assign same input for states belonging to a single output group.
\end{itemize}
Follow this control law (output feedback) till state estimate converges; there-after follow state feedback.

\subsection*{Advantages:}
\begin{enumerate}
    \item This could be, depending upon requirements, computationally much more efficient
    \item Observer can be constructed even for the systems non-reconstrctible in general sense.
\end{enumerate}
The second point follows from the fact that any system, if follows distinguishable trajectory till the state is accurately reconstructed, then from the next evolution onward the system will remain `observed' i.e. the system state will always be known (provided no disturbances/ faults/ uncertainties are present).

\subsection*{Further Work:}
If the control is provided through a different $BN$, refered to as control $BN$, then analysis based on the following line of reasoning may be implemented:
\begin{enumerate}
    \item Find all the attractors (cycles and fixed points) of the control $BN$
    \item Starting from all the states of the (primary) $BCN$ as initial states, check for only the $(x,u)$ trajectories with control sequence as the attractors of the control $BN$ (secondary).
\end{enumerate}

\begin{figure}[t]
\centering
\includegraphics[scale=1]{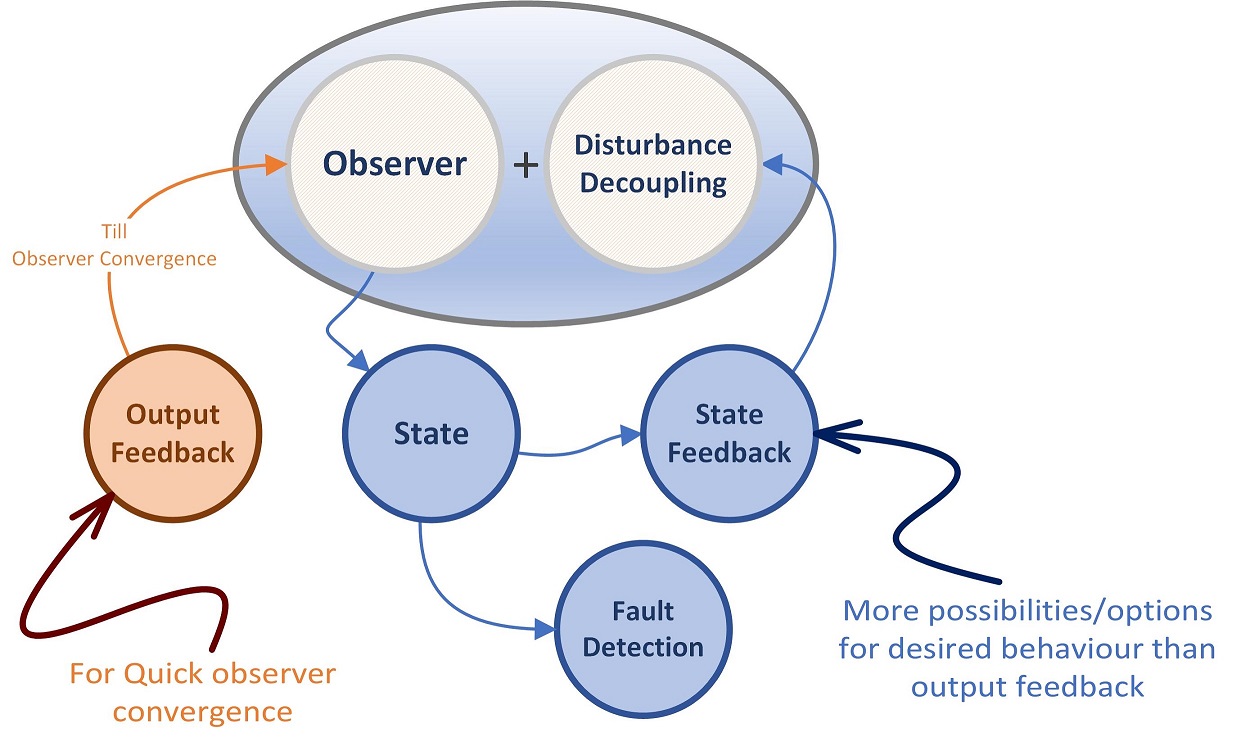}
\caption{Output feedback application}
\label{output_feedback_observer_P_DD}
\end{figure}

\section{Instantaneous Fault Detection}
Detection of fault is, in general, a tricky process which requires analysis of a sequence of inputs and corresponding outputs along with the system dynamics to provide viable result/conclusion on whether a fault has occurred or not. Results on fault detection in $BCN$ available are based on similar approach. If the underlying system satisfies a few extra conditions, then the fault can be detected instantaneously(in the very next instant after the fault has occurred). This needs for reconstructibility to hold true, as presented in \cite{fornasini2015fault} and requires an extra condition in the form of an invertible map between fault and output, i.e. if $H_f:(fault)\mapsto (output)$ is a map fault and output keeping input and state constant, then $H_f$ is invertible/pseudo-invertible \{see for reference \cite{zhang2015invertibility}\}(non-singular). This can be justified by arguing that that, in general case a fault is detected when a map $H_u:(u_1,\dots ,u_k)\mapsto (y_1,\dots ,y_k)$ for some integer $k\geq 1$ produces a sequence of outputs $(\tilde{y}_1,\dots ,\tilde{y}_j)$ for a sequence of inputs $(\bar{u}_1,\dots ,\bar{u}_j)$ that is different from/than the expected unique set of outputs $(\bar{y}_1,\dots ,\bar{y}_j)$. The fault can be instantaneously detected if $H_u$ under the effect of fault produces an output, different than the expected unique output, for $k=1\ \forall u_i$. 

This requires every fault to alter the output when it is present, which can be formalised by an invertible $H_f$ when $(the\ number\ of\ outputs) = (the\ number\ of\ faults)$ and a pseudo-invertible $H_f$ when $(the\ number\ of\ outputs) \geq (the\ number\ of\ faults)$. This can be summarised as:
\begin{itemize}
    \item Appearance of fault should cause changes in the output
    \item Changes should be unique for each fault combination
\end{itemize}

\{Smaller output friendly set and smaller number of outputs was helpful in disturbance decoupling, but opposite is true for fault detection.\}
\begin{itemize}
    \item Observability 
        \begin{itemize}
            \item Identification of fault from sequence of input-state-output \underline{OR} Equivalent (maybe).
        \end{itemize}
    \item Reflectivity, Reflective variables (Full rank condition)
        \begin{itemize}
            \item Output $\rightarrow$ Fault Inverse Map (Non-singularity)
        \end{itemize}
    \item Fault detection: necessary $(Full\ Rank)$ and sufficient $(Under\ Nonsigular\ F \rightarrow Y\ Mapping)$ condition
\end{itemize}

\subsection{Observer Aided Instantaneous Fault Detection in BCN}
In any system detection and isolation of fault has significant importance. Faults may be treated as uncertainties that deteriorate the performance of the system. The faults are the result of the internal system malfunction affecting the performance for all future times till actively removed. The effect of the fault especially in the critical systems, could be devastating. In $BCN$ the disturbances and the faults are represented in the linear form in the exact same manner with a difference in their characterization. The disturbance is expected to pertain for only a few steps of system evolution. On the other hand the fault remains till it is actively addressed. Hence the strategies to deal with the disturbance is to avoid its effect, but for the fault it is to detect and act on it.
 
To detect the fault instantaneously the following condition must hold:\\
For every output set $O_{si}$ there should exist some output set $O_{sj}\  i,j\in \{1,2,\dots,2^p\}$  such that under the normal system dynamics no state from $O_{si}$ transitions to $O_{sj}$ directly i.e. $\nexists x_j \in O_{si}$ such that for any $x_i\in O_{si}\    Lx_i=x_j$; in other words output sequence $y_i\rightarrow y_j$ is impossible. Therefore if the output sequence $y_i\rightarrow y_j$ is observed then the fault has occurred.

\subsubsection*{Structure required to detect fault instantaneously}
\begin{itemize}
    \item Every sub-block is full ranked.
    \item Element corresponding to fault in every sub-block $(i)$ belongs to the impossible output set for $O_{si}$ i.e. $I_m(O_{si})$.
    \item For multiple faults there need to be at-least as many number of impossible output sets for $O_{si}$
\end{itemize}
\subsubsection*{The states are available via observer}
In this case conditions presented earlier can be relaxed to some extent
\begin{itemize}
    \item Every sub-block is full ranked.
    \item Elements in every sub-block$(i)$ belong to different output groups
    \item For multiple faults there need to be at-least as many different outputs and every element of sub-block$(i)$ belong to different output groups.
\end{itemize}
Assuming that,
\begin{enumerate}
    \item Some output feedback exits that aids (drives) observer (i.e. observer aiding output feedback)
    \item System remains fault free at-least till observer converges.
\end{enumerate}
Therefore if possible, instantaneous fault detectable $BN$ can be constructed using state feedback from $BCN$ with the help of observer aiding O/P feedback.

\begin{figure}[t]
\centering
\includegraphics[scale=0.95]{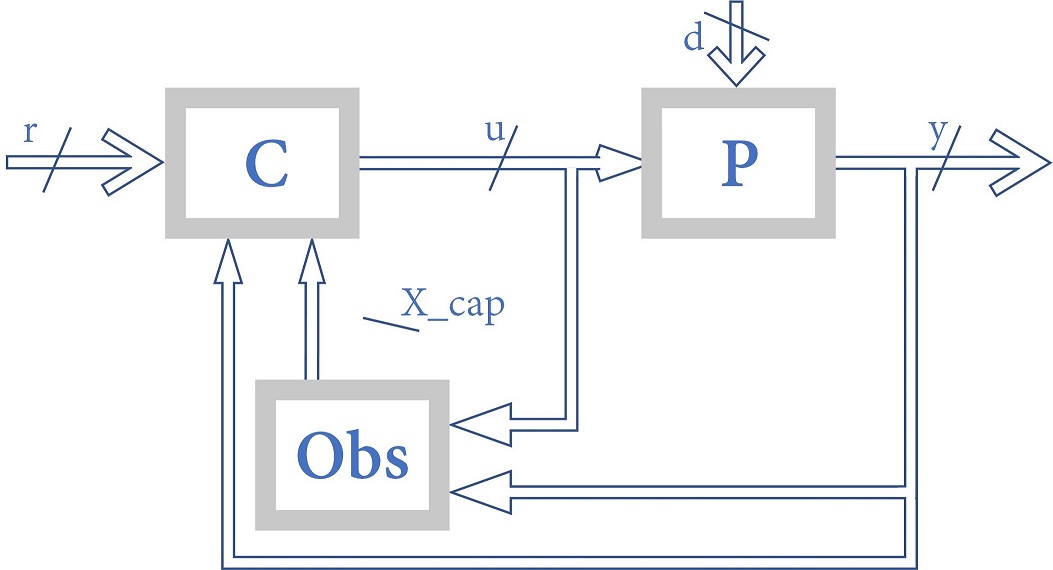}
\caption{Observer aided fault detection: Reference tracking}
\label{Observer_output_feedback_reference_tracking}
\end{figure}

\subsection{\textbf{Algorithm} Instantaneous Fault Detection:}
Let the BCN dynamics be given by
$$x_y^+=Lux\xi_f$$
where, $x_y$ is $y$ friendly $d$-dimensional sub-system, $x$ is $n$ dimensional complete state, $u$ is $m$-dimensional input and $\xi_f$ is $t$-dimensional fault vector. An algorithm to check for the possibility of instantaneous fault detection is:

\begin{algorithm}
\caption{Instantaneous Fault Detection}\label{alg:Fault_Inst}

\begin{enumerate}
    \item Divide $L$ matrix into $2^m$ blocks $L_u^1$ to $L_u^{2^m}$ of size $(2^{n+t}\times 2^{n+t})$
    \item Divide every block $L_u^i$ for $i\in \{1,\dots ,2^m\}$ into $2^n$ sub-blocks $L_{uy}^{i1}$ to $L_{uy}^{i2^n}$ of size $(2^{d}\times 2^{t})$
    \item For fault detection
        \begin{itemize}
            \item At-least one $j^{th}$ sub-block should rank $2^t$
            \item The map $F_m:t\mapsto y$ should be invertible or pseudo invertible
        \end{itemize}
    \item Construct sets $C_j,\ \forall i\in \{1,\dots ,2^n\}$ as
    $$C_j:=\{i\ |\ L_{uy}^{ij}\ is\ full-ranked\}$$
\end{enumerate}

\end{algorithm}

The state feedback controller is selected as
$$M_x=\delta_{2^m}[C_1^{i_1}\ C_2^{i_2}\ \dots\ C_{2^n}^{i_{2^n}}]$$
where $C_j^{ij}\in \mathcal{L}_{2^m\times 2^{n-d}}$ and $Col(C_j^{ij})\in C_j$. Total number of such controllers possible are:
$$N_{tc}=|C_1|\cdot |C_2|\dots |C_{2^n}|$$
where $|\cdot|$ indicates cardinality of a set.

\textbf{E.g. 11.}
$$\begin{matrix}
L= & \delta_8 [\ 1\ 2\ 1\ 2\ 3\ 4\ 3\ 4\ 1\ 1\ 1\ 1\ 3\ 1\ 3\ 3\\
    & \; 2\ 3\ 2\ 2\ 4\ 4\ 4\ 4\ 1\ 1\ 1\ 2\ 3\ 3\ 3\ 3\\ 
    & \; 1\ 4\ 1\ 2\ 3\ 4\ 3\ 4\ 1\ 3\ 1\ 1\ 3\ 3\ 4\ 3\\ 
    & \; 2\ 4\ 2\ 2\ 4\ 4\ 4\ 4\ 1\ 3\ 1\ 1\ 3\ 3\ 3\ 4\ ]
\end{matrix}$$
The output friendly sub-system has the form $x_y^+=Lux\xi_f$. Then applying the \textit{Algorithm B} various sets are obtained as
$$C_1=\{1,2,3,4\},\ C_2=\{1,3\},\ C_3=\{1,3\},\ C_4=\{1,3\}$$
$$C_5=\{3,4\},\ C_6=\{2\},\ C_7=\{1\},\ C_8=\{3,4\}$$
This results in a total of $4\cdot 2\cdot 2\cdot 2\cdot 2\cdot 1\cdot 1\cdot 2=128$ state feedback matrices that instantaneously detect the fault if occurred in the system, a possible one of which can be given as
$$M_x=\delta_4[4\ 3\ 3\ 3\ 4\ 2\ 1\ 4]$$
$$y^+=Hx^+=HLM_x\psi_nx\xi_f$$
$2^n$ blocks of $2^f$ columns each $:\rightarrow$ each block full ranked\\
\subsection{Reflective Variable:}
Let $G:\mathcal{D}^n\mapsto \mathcal{D}^k$ be a multi-variable Boolean function. A Boolean variable is said to be reflective if
\begin{multline}
    g_j(x_1,\dots ,x_{i-1},1,x_{i+1},\dots ,x_n)\\
    \neq g_j(x_1,\dots ,x_{i-1},0,x_{i+1},\dots ,x_n)
\end{multline}
for all values of $x_k\in \{1,\dots,2^n\}/i$ for all $j\in \{1,\dots ,k\}$.

\textbf{Lemma 14 (Fault):} Let $M_G\in \mathcal{L}_{2^k\times 2^n}$ be the structure matrix of the logical mapping $G:\mathcal{D}^n\mapsto \mathcal{D}^k$ and let an integer, $0 < r\leq n$, be given. Split $M_G$ into $2^r$ blocks as $M_G=[M_1,M_2,\dots,M_{2^r}]$ where, $M_1,\dots,M_{2^r}\in \mathcal{L}_{2^k\times 2^{n-r}}$. Then $x_{r+1},\dots ,x_n$ are all reflective variables iff, rank $M_i=(2^{n-r})\ \forall i\in \{1,\dots ,2^r\}$, i.e. $M_i$ is full column ranked.

\textbf{Proof:}\hfill ($\Rightarrow$ Necessary)\\
Suppose $\ltimes_{j=1}^{r}x_j=\delta_{2^r}^i$. Then $M_G\ltimes_{j=1}^{r}x_j=M_i$. Since,
$$\ltimes_{j=1}^{k} g_j(x_1,\dots,x_n)=M_G\ltimes_{j=1}^{n}x_j=M_i\ltimes_{j=r+1}^{n}x_j$$ 
and the variables, $x_{r+1},\dots, x_n$ are reflective, it follows that all different values of $\ltimes_{j=r+1}^{n}x_j$ will produce different $M_i\ltimes_{j=r+1}^{n}x_j$ vector. Therefore all columns of $M_i$ are distinct and hence $rank(M_i)=2^{n-r}$.

\hfill($\Leftarrow$ Sufficient)\\
Since rank$(M_i)=2^{n-r}$ and $M_i\in \mathcal{L}_{2^k\times 2^{n-m}}$, all columns of $M_i$ are different, therefore $M_I\ltimes_{j=r+1}^{n}x_j$ is a unique vector for every $x_j$.
$$\ltimes_{j=1}^{k} g_j(x_1,\dots,x_n)=M_i\ltimes_{j=r+1}^{n}x_j$$
it follows that $\ltimes_{j=1}^{k} g_j(x_1,\dots,x_n)$, is also unique for distinct values of $x_{r+1},\dots,x_n$ and therefore the variables $x_{r+1},\dots,x_n$ are reflective.

\hfill(\textbf{Condition F})
\subsection{\textbf{Theorem} Instantaneous Fault Detection:}
\begin{eqnarray*}
    x^+=Lux\xi_f\\
    y=Hx\\
    y^+=HLux\xi_f
\end{eqnarray*}
$u:m-$ inputs, $x:n-$ states, $y:p-$ outputs, $\xi_f:f-$ faults.

\textbf{Problem Statement:} Identifying fault/s with state feedback $(M_x)$
\begin{eqnarray*}
    x^+=LM_x\psi_nx\xi_f\\
    y^+=HLM_x\psi_nx\xi_f
\end{eqnarray*}

\textbf{Theorem 15:} In the BCN described by above mentioned dynamics, fault/s can be detected instantaneously if every fault is reflected in output separately, i.e. $M^O$ satisfies \textit{Lemma.(fault)} for every element of $\xi_f$.

\textbf{Proof:}\hfill ($\Rightarrow$ Necessity)\\
Let $\tilde{L}=LM_x\psi_n$, then $\tilde{L}$ has the following structure
\begin{multline}
    \tilde{L}=[Blk(2^{i_1}+1)\ of\ L\ |\ Blk(2^{i_2}+2)\ of\ L \\
    |\dots\ |\ Blk(2^{i_{2^n}}+2^n)\ of\ L ]
\end{multline}
where $i_k\in \{0,\dots,2^n-1\}$. Since, $M^O$ satisfies \textit{Lemma(fault)}, it can be divided into $2^n$ blocks of size $(2^p\times 2^f)$, each with full column rank.\\
Let $Col(H)^r:=\{j\ |\ Col_j(H)=y_r\}$; $y_r-r^{th}$ output. Then, $Col(H)^r$ indicates set of column indices in $r^{th}$ row of $H$, where matrix entry is 1. $(M^O)_{ij}=1$ iff $[Col_j(\tilde{L})]_m=1$ and $m\in Col(H)^i$\\
The columns of each su-block should belong to different $y^r$. If any two columns are from same output group then the entry `1' for these two columns will be in same row $i$ and this will violet the \textbf{Lemma(fault)}. The \textbf{Lemma(fault)} is satisfied iff, for
\begin{multline}
    e_o[Col_{j_1}(\tilde{L})]\in Col(H)^{i_{j_1}} \ \&\ e_o[Col_{j_2}(\tilde{L})]\in Col(H)^{i_{j_2}};\\
    Col_{j_1}(\tilde{L}),Col_{j_2}(\tilde{L})\in [Blk_k\tilde{L}] \ \&\ i_{j_1}\neq i_{j_2}
\end{multline}
where $e_o(\mathcal{V}_B):=\{l\ |\ \mathcal{V}_{Bl}=1\}$, is the entry `1' element of Boolean vector $\mathcal{V}_B$. By (23) and (24) it is clear that $L$ needs to satisfy the given condition.

\hfill ($\Leftarrow$ Sufficiency)

Since, $L$ matrix satisfies given conditions, it is possible to construct $\tilde{L}$ with appropriate feedback matrix $M_x$ using algorithm \ref{alg:Fault_Inst} \hfill $\square$

By (23), all the $2^n$ blocks of $\tilde{L}$ will have all columns belonging to different output groups. Simple matrix multiplication indicates that for each $(2^p\times 2^f)$ sized block of $M^O$, the entry `1' will be different rows.


\section{Disturbance Decoupling + Instantaneous Fault Detection}
Disturbance decoupling ($DD$) requires intervention through input to nullify the effect of disturbance. Fault detection ($FD$) on the other hand is based on observation of input and output sequence of a sufficient length to reconstruct the state to detect the fault, rather than avoiding the faulty state. Therefore, the importance of the feedback differs for both the cases. In presence of state feedback, the problem of Fault detection is trivial, since state is available for observation (hence no need for reconstruction). Therefore, the $DD + FD$ problem with feedback is simply the problem of $DD$ with state feedback; which can be with any of the available methods (including the ones presented here). Assuming availability of the output feedback with some special property, the following three possibilities arise:
\begin{enumerate}
    \item \textit{Neither fault nor disturbance appear till the state is reconstructed:}\\
    In this case an output feedback that aids the observer (or doesn't affect the observer adversely) will suffice 
    \item \textit{Only disturbance appears till the state is reconstructed:}\\
    In this case output feedback that is capable of $DD$ and aids (doesn't oppose) the observer is required
    \item \textit{Both disturbance and fault appear before the state is reconstructed:}\\
    Neither of the objective can be achieved with guarantee
\end{enumerate}
The fault and the disturbance can be incorporated in the system dynamics with one of the following two ways:
\begin{eqnarray}
 \label{eq_fault_P_distur_linear} X^+ = L_1ux\xi_f\xi_d\\ \label{eq_distur_P_fault_linear} X^+ = L_2ux\xi_d\xi_f
\end{eqnarray}
where, $\xi_f$ is the Boolean vector of fault and $\xi_d$ is the Boolean vector of disturbance.
\subsubsection*{\textbf{i}} Dividing $L_1$ (state transition matrix from equation (\ref{eq_fault_P_distur_linear})) into $2^{n+m}$ blocks, each block will have the following structure:
\begin{center}
\begin{tabular}{ |c|c|c|c| } 
 \hline
 nf-nd & nf-d & f-nd & f-d\\ 
 \hline
\end{tabular}
\end{center}
with entries of the block indicating the next (future) state of the system when, 
\begin{itemize}
    \item $nf-nd:$ no fault and no disturbance is present
    \item $nf-d:$ no fault, but disturbance is present
    \item $f-nd:$ no disturbance but fault is present
    \item $f-d:$ both fault and disturbance are present
\end{itemize}
Then,\\
$\mathbf{DD}$ is possible if under some input:\\ 
\textit{(no fault)} $\left | nf-nd \right |$ and $\left | nf-d \right |$ blocks are same.\\
\textit{(with fault)} $\left | f-d \right |$ and $\left | f-nd \right |$ blocks are same.\\
$\mathbf{FD}$ is possible if under some input:\\ 
$\left | nd-nf \right |$ and $\left | nd-f \right |$ blocks belong to different output group if state is known (can be observed).\\
\centerline{OR}\\
$\left | nd-f \right |$ block belongs to impossible output group if state is not known (can not be observed).
\subsubsection*{\textbf{ii}} Dividing $L_2$ (state transition matrix from equation (\ref{eq_distur_P_fault_linear})) into $2^{n+m}$ blocks, each block will have the following structure:
\begin{center}
\begin{tabular}{ |c|c|c|c| } 
 \hline
 nd-nf & nd-f & d-nf & d-f\\ 
 \hline
\end{tabular}
\end{center}
Then,\\
$\mathbf{DD}$ is possible if under some input:\\ 
\textit{(no fault)} $\left | nd-nf \right |$ and $\left | d-nf \right |$ blocks are identical.\\
\textit{(with fault)} $\left | nd-f \right |$ and $\left | d-f \right |$ blocks are identical.\\
$\mathbf{FD}$ is possible if under some input:\\ 
$\left | d-nf \right |$ and $\left | d-f \right |$ blocks belong to different output group if state is known (can be observed).\\
\centerline{OR}\\
$\left | d-f \right |$ block belongs to impossible output group  if state is not known (can not be observed).

These results can be summarised as:\\
\textbf{Observer:} No $\xi_f$ till observer convergence:\\
$\Rightarrow \underset{DD}{\underbrace{\textit{Output f/b till observer convergence + Identical Blocks}}}$\\
\centerline{$\Rightarrow$ State feedback after observer convergence}\\  
\centerline{$\Rightarrow \underset{DD}{\underbrace{\textit{Identical Blocks}}} + \underset{FD}{\underbrace{\textit{Distinct  Outputs}}}$}\\
\textbf{No Observer:} (Instantaneous FD) Output feedback:\\
\centerline{$\Rightarrow \underset{DD}{\underbrace{\textit{Identical Blocks}}} + \underset{FD}{\underbrace{\textit{Impossible Output Group}}}$}\\
Observer can work as fault detector in absence of $\xi_d$, if instantaneous $FD$ is not possible.\footnote{If the set of possible states that correspond to tuple ($y_c,u_c,Y^+$) is empty $\Rightarrow $ fault} \footnote{In presence of disturbance, all possibilities with disturbance as an unknown input may be considered to form the set of all possible states corresponding to ($y_c,u_c,y^+$) tuple}\\
It can be observed that, the objectives of $DD$ and $FD$ can be achieved with certainty only if the system state is reconstructed instantaneously (observer converges instantaneously). Consequently, in this case fault can be detected instantaneously. Therefore, the problem of solving for $DD + FD$ with certainty is equivalent to the problem of solving for $DD + IFD$ (instantaneou $FD$).

\subsection{\textbf{Algorithm} Instantaneous Fault Detection and Disturbance Decoupling:}
Let the BCN dynamics be given by
$$x_y^+=Lux_yx_r\xi_d\xi_f$$
where, $x_y$ is $y$ friendly $d$-dimensional sub-system, $x_r$ is $n-d$-dimensional vector of remaining states, $u$ is $m$-dimensional input, $\xi_d$ is $s$-dimensional disturbance vector and $\xi_f$ is $t$-dimensional fault vector. Then the algorithm for DD with instantaneous fault detection is given by:

\begin{algorithm}
\caption{Disturbance Decoupling DD with Instantaneous Fault Detection}\label{alg:DD_Inst_Flt_Dtec}

\begin{enumerate}
    \item Divide $L$ matrix into $2^m$ blocks $L_u^1$ to $L_u^{2^m}$ of size $(2^{n+s+t}\times 2^{n+s+t})$
    \item Divide every block $L_u^i$ for $i\in \{1,\dots ,2^m\}$ into $2^n$ sub-blocks $L_{uy}^{i1}$ to $L_{uy}^{i2^n}$ of size $(2^{d}\times 2^{s+t})$
    \item For every sub-block index $j$ there needs to be a sub-block, corresponding to some $u$, such that the corresponding sub-block $j$ is structured in the following way\\
    For 1-Disturbance and 1-Fault:
        \begin{itemize}
            \item $\delta_{2^d}^k=\delta_{2^d}^l$ if $k\ \&\ l$ are both either even or odd
            \item $\delta_{2^d}^k\neq \delta_{2^d}^l$ if one of $k\ \&\ l$ is even the other is odd
        \end{itemize}
    For s-Disturbances and t-Faults:\\
    Make $2^s$ equal divisions of each sub-block $j$, then if
        \begin{itemize}
            \item all divisions are identical
            \item all divisions have full rank
            \item the $F_m:t\mapsto y$ map is invertible or pseudo-invertible
        \end{itemize}
\end{enumerate}

\end{algorithm}

If the conditions in the algorithm are satisfied, then a single controller can be utilized to detect faults instantaneously as well as to decouple the disturbance signal.

The state feedback controller is selected as
$$M_x=\delta_{2^m}[C_1^{i_1}\ C_2^{i_2}\ \dots\ C_{2^n}^{i_{2^n}}]$$
where $C_j^{ij}\in \mathcal{L}_{2^m\times 2^{n-d}}$ and $Col(C_j^{ij})\in C_j$. Total number of such controllers possible are:
$$N_{tc}=|C_1|\cdot |C_2|\dots |C_{2^n}|$$
where $|\cdot|$ indicates cardinality of a set.

\textbf{E.g. 12.}
$$\begin{matrix}
L= & \delta_8 [\ 1\ 2\ 1\ 2\ 3\ 4\ 3\ 4\ 1\ 1\ 1\ 1\ 3\ 3\ 3\ 3\\ 
    & \; 2\ 3\ 2\ 2\ 4\ 4\ 4\ 4\ 1\ 1\ 1\ 1\ 3\ 3\ 3\ 3\\
    & \; 1\ 4\ 1\ 2\ 3\ 4\ 3\ 4\ 1\ 3\ 1\ 3\ 3\ 3\ 3\ 3\\
    & \; 2\ 4\ 2\ 2\ 4\ 4\ 4\ 4\ 1\ 3\ 1\ 1\ 2\ 4\ 2\ 4\ ]
\end{matrix}$$
The output friendly sub-system has the form $x_y^+=Lux\xi_d\xi_f$. Then applying the \textit{Algorithm C} various sets are obtained as
$$C_1=\{1\},\ C_2=\{1,3\},\ C_3=\{3\},\ C_4=\{4\}$$
This results in a total of $1\cdot 2\cdot 1\cdot 1=2$ state feedback matrices that instantaneously detect the fault if occurred in the system, which are be given as
$$M_x^1=\delta_4[1\ 1\ 3\ 4]$$
$$M_x^1=\delta_4[1\ 3\ 3\ 4]$$
\subsection{Redundant Variables + Reflective Variables:}
\textbf{Lemma 16 (Fault + Disturbance):} Let $M_G\in \mathcal{L}_{2^k\times 2^n}$ be the structure matrix of the logical mapping $G:\mathcal{D}^n\mapsto \mathcal{D}^k$, and let an integer $r\leq n$. Split $M_G$ into $2^n$ blocks as, $M_G=[M_1, M_2, \dots, M_{2^r}]$ where, $M_1, M_2, \dots, M_{2^r}\in \mathcal{L}_{2^k\times 2^{n-r}}$. Let, $0<s\leq n-r$, be an integer. Split $M_i$ into $2^s$ blocks as, $M_i=[M_i^1, M_i^2, \dots, M_i^{2^s}]$. Then $x_{r+1}, \dots, x_{r+s}$ are all redundant variables and $x_{r+s+1}, \dots, x_{n}$ are all reflective variables iff rank$(M_i^j)=2^{n-r-s}$ and $M_i^j=M_i^k\ \forall\ j,k\in \{1,\dots,2^s\}$.

\textbf{Proof:} \hfill ($\Rightarrow$ Necessity)\\
Suppose $\ltimes_{j=1}^{r}x_j=\delta_{2^r}^i$, then $M_G\ltimes_{j=1}^{r}x_j=M_i$. Let $\ltimes_{j=r+1}^{s}x_j=\delta_{2^s}^k$, then $M_i\ltimes_{j=r+1}^{s}x_j=M_i^k$.\\
Since,
\begin{multline}
    \ltimes_{j=1}^{k} g_j(x_1,\dots,x_n)=M_G\ltimes_{j=1}^{n}x_j\\
    =M_i\ltimes_{j=r+1}^{n}x_j=M_i^k\ltimes_{j=s+1}^{n}x_j
\end{multline}
variables $x_{r+1},\dots, x_s$ are redundant and $M_i\ltimes_{j=r+1}^{r+s}x_j=M_i^k$; it follows that $M_i^{k_1}=M_i^{k_2}\ \forall\ k_1,k_2\in \{1,\dots, 2^s\}$ i.e. $M_i\ltimes_{j=r+1}^{r+s}x_j$ is a fixed matrix, therefore all the blocks of $M_i$ are identical. Variables $x_{s+1}, \dots, x_{n}$ are reflective and $\ltimes_{j=1}^{k} g_j(x_1,\dots,x_n) = M_i\ltimes_{j=r+1}^{n}x_j$, it follows that $M_i^k\ltimes_{j=s+1}^{n}x_j$ produces distinct vectors for distinct $\ltimes_{j=s+1}^{n}x_j$. Therefore all columns of $M_i^k$ are distinct and rank$(M_i)=2^{n-r-s}$.

\hfill($\Leftarrow$ Sufficiency)\\
Since, rank$(M_i^k)=2^{n-r-s}$ and $M_i^{k_1}=M_i^{k_2}\ \forall\ k_1,k_2\in \{1,\dots,2^s\}$, where $M_i^k\in \mathcal{L}_{2^k\times 2^{n-r-s}}$, all columns of $M_i^k$ are distinct, therefore $M_i^k\ltimes_{r+s+1}^{n}x_j$ is a unique for vector every $\ltimes_{r+s+1}^{n}x_j$ and $M_i^k\ltimes_{r+1}^{r+s}x_j$ is identical for any vector $\ltimes_{r+1}^{r+s}x_j$ with 
$$\ltimes_{j=1}^{k} g_j(x_1,\dots,x_n)=M_i\ltimes_{j=r+1}^{r+s}x_j \ltimes_{l=r+s+1}^{n}x_l$$
it follows that, $\ltimes_{j=1}^{k} g_j(x_1,\dots,x_r,x_{r+1},\dots,x_{r+s},\dots,x_n)$ is same for any combination of values of $x_{r+1},\dots,x_{r+s}$ and is unique for every combination of values of $x_{r+s+1},\dots,x_{n}$. \hfill $\square$
\subsection{\textbf{Theorem} Disturbance Decoupling and Instantaneous Fault Detection :} 
\begin{eqnarray*}
    x^+ &=& Lux\xi_d\xi_f\\
    y^+ &=& HLux\xi_d\xi_f
\end{eqnarray*}
$u:m-$ inputs, $x:n-$ states, $y:p-$ outputs, $\xi_d:d-$ disturbances, $\xi_f:f-$ faults.

\textbf{Problem Statement:} Identifying fault and decoupling disturbance with state feedback $(M_x)$
\begin{eqnarray*}
     y^+=HLM_x\psi_nx\xi_d\xi_f=M^Ox\xi_d\xi_f
\end{eqnarray*}

\textbf{Theorem 17:} In the BCN described by above mentioned dynamics, instantaneous fault detection and disturbance decoupling can be achieved if every fault variable is reflected in output separately and every disturbance variable is redundant, i.e. $M^O$ satisfies lemma (Fault + Disturbance) for every element of $\xi_f$ and $\xi_d$.

\textbf{Proof:} \hfill ($\Leftarrow$ Necessity)\\
Let $\tilde{L}=LM_x\psi_n$. $\tilde{L}$ has following structure
\begin{multline}
    \tilde{L}=[Blk(2^{i_1}+1)\ of\ L\ |\ Blk(2^{i_2}+2)\ of\ L \\
    |\dots\ |\ Blk(2^{i_{2^n}}+2^n)\ of\ L ]
    \label{eq:STM_FD}
\end{multline}
where $i_k\in \{0,\dots,2^n-1\}$. Since, $M^O$ satisfies \textit{Lemma(fault)}, it can be divided into $2^n$ blocks of size $(2^p\times 2^f)$, each with full column rank.\\
Let $Col(H)^r:=\{j\ |\ Col_j(H)=y_r\}$; $y_r-r^{th}$ output. Then, $Col(H)^r$ indicates set of column indices in $r^{th}$ row of $H$, where matrix entry is 1. 
\begin{multline}
    (M^O)_{ij}=1 \text{ iff } [Col_j(\tilde{L})]_m=1 \ \&\  m\in Col(H)^i
\end{multline}
For all the $2^d$ sub-blocks of $[Blk_j\tilde{L}]$ ot construct $M^O$ that satisfies lemma (Fault + Disturbance) for a given $H$, it can be observed that, for $[Blk_j\tilde{L}]_k$ (i.e. $k^{th}$ sub-block of $[Blk_j\tilde{L}]$), if 
\begin{eqnarray*}
    Col_{l_1}([Blk_j\tilde{L}]_k)\in y_{r_{l_1}}\\
    Col_{l_2}([Blk_j\tilde{L}]_k)\in y_{r_{l_2}}\\
    e_o[Col_{l_1}([Blk_j\tilde{L}]_k)]\in Col(H)^{i_{l_1}}\\
    e_o[Col_{l_2}([Blk_j\tilde{L}]_k)]\in Col(H)^{i_{l_2}}
\end{eqnarray*}
then,
\begin{multline}
     y_{r_{l_1}}\neq y_{r_{l_2}}\ \forall l_1,l_2\in Col([Blk_j\tilde{L}]_k)\\
    Col(H)^{i_{l_1}}\neq Col(H)^{i_{l_2}}\ \forall l_1,l_2\in Col([Blk_j\tilde{L}]_k)
    \label{eq:impossible_op}
\end{multline}
   
and if,
\begin{eqnarray*}
    Col_{l}([Blk_{j_1}\tilde{L}]_k)\in y_{r_{j_1}}\\
    Col_{l}([Blk_{j_2}\tilde{L}]_k)\in y_{r_{j_2}}\\
    e_o[Col_{l}([Blk_{j_1}\tilde{L}]_k)]\in Col(H)^{i_{j_1}}\\
    e_o[Col_{l}([Blk_{j_2}\tilde{L}]_k)]\in Col(H)^{i_{j_2}}
\end{eqnarray*}
then,
\begin{multline}
    y_{r_{j_1}}\neq y_{r_{j_2}}\ \forall j_1,j_2\in \{1,\dots,2^d\}\\
    Col(H)^{i_{j_1}}\neq Col(H)^{i_{j_2}}\ \forall j_1,j_2\in \{1,\dots,2^d\}
    \label{eq:impossible_op_all_Disturbances}
\end{multline}
    
From (\ref{eq:STM_FD}), (\ref{eq:impossible_op}) and (\ref{eq:impossible_op_all_Disturbances}) it is clear that all the conditions must be satisfied.

\hfill ($\Leftarrow$ Sufficiency)\\
Since, $L$ satisfies given condition, it is possible to construct $\tilde{L}$ with appropriate feedback matrix $M_x$ using algorithm \ref{alg:DD_Inst_Flt_Dtec}\\
(\ref{eq:STM_FD}) and simple matrix multiplication indicate that, resulting $H\tilde{L}$ matrix will satisfy lemma (Fauld + Disturbance). \hfill $\square$
\newline



It can be verified that for instantaneous fault detection with disturbance decoupling the matrix $LM_x\psi_n$ (with $M_x$ constructed according to algorithm) makes system insensitive to noise and satisfies the necessary and sufficient conditions of theorem 3.9 from \cite{zhang2015invertibility}, therefore the system is fault-output invertible, hence occurrence of fault can be uniquely identified from the output of the system.

\appendices
\section{}
\subsection{\textbf{Algorithm} Disturbance decoupling in finite iterations:}
Let the BCN dynamics be given by
$$x^+=Lux_yx_r\xi_d$$
where, $x_y-$ is $y$ friendly $d$-dimensional sub-system, $x_r$ is remaining $n-d$ dimensional sub-system, $u$ is $m$-dimensional input and $\xi_d$ is $s$-dimensional disturbance. Then the algorithm for DD in Finite Iterations is given by:

\begin{algorithm}
\caption{Disturbance Decoupling in Finite Iteration: 2}\label{alg:DD_Fin_Iter_2}

\begin{enumerate}
    \item Divide $L$ matrix into $2^m$ blocks $L_u^1$ to $L_u^{2^m}$ of size $(2^{n+s}\times 2^{n+s})$
    \item Check for all the sub-blocks corresponding to indices in $(S_1)$, for every $\delta_{2^d}^i$,if $i\in(S_1)$ 
    \item Divide every block $L_u^i$ for $i\in \{1,\dots ,2^m\}$ into $2^d$ sub-blocks $L_{uy}^{i1}$ to $L_{uy}^{i2^d}$ of size $(2^{n-d+s}\times 2^{n-d+s})$
    \item 
        \begin{enumerate}
            \item Construct a set of sub-block indices $(S_1)$, such that the corresponding sub-block has rank 1 at least in one block.
            \item Construct a set of indices $(S_2)$, such that for the corresponding sub-blocks, for every column $\delta_{2^d}^k$, $k\in (S_1)$.
            \item Construct a set of sub-block indices $(S_3)$, such that for the corresponding sub-blocks, for every column $\delta_{2^d}^k$, $k\in (S_1)\bigcup (S_2)$.
            \item \dots
        \end{enumerate}
    Construct sets $(S_i)$ until all states/indices are classified or the remaining states/indices can not be classified
    \item Construct sets $C_j,\ \forall i\in \{1,\dots ,2^d\}$ as
    $$C_j:=\{i|L_{uy}^{ij}\ s.t\ if\ j\in(S_l)\ then\ CS_j\in \bigcup_{i=1}^{j-1}S_i\}$$
    where $CS_j:=\{Column\ set\ of\ L_{uy}^{ij}\}$ 
\end{enumerate}

\end{algorithm}

Then the disturbance can be decoupled in at-most $j$ iterations, if all the indices are classified and condition $2)$ is met; where $j$ is the largest integer such that $S_j\neq \{\phi \}$ (Or the number of sets returned by the algorithm).

\textit{Controller Selection:}\\
For every sub-block index $j$, the input is chosen from any of the block indices which transition $y$-friendly sub system from $S_l$ to $S_{l-1}$. This is implemented in step $5)$ by constructing sets $C_1$ to $C_{2^d}$. Then the state feedback controller is selected as
$$M_x=\delta_{2^m}[C_1^{i_1}\ C_2^{i_2}\ \dots\ C_{2^d}^{i_{2^d}}]$$
where $C_j^{ij}\in \mathcal{L}_{2^m\times 2^{n-d}}$ and $Col(C_j^{ij})\in C_j$. Total number of such controllers possible are:
$$N_{tc}=|C_1|^{2^{n-d}}\cdot |C_2|^{2^{n-d}}\dots |C_{2^d}|^{2^{n-d}}$$
where $|\cdot|$ indicates cardinality of a set.

\textbf{E.g. 13.}
$$\begin{matrix}
L= & \delta_8 [\ 1\ 2\ 1\ 2\ 3\ 4\ 3\ 4\ 2\ 2\ 2\ 2\ 3\ 3\ 3\ 3\\ 
    & \; 2\ 3\ 2\ 2\ 4\ 4\ 4\ 4\ 1\ 1\ 1\ 1\ 3\ 3\ 3\ 3\\
    & \; 1\ 4\ 1\ 2\ 3\ 4\ 3\ 4\ 1\ 3\ 1\ 1\ 3\ 3\ 3\ 3\\ 
    & \; 2\ 4\ 2\ 2\ 4\ 4\ 4\ 4\ 1\ 3\ 1\ 1\ 3\ 3\ 3\ 3\ ]
\end{matrix}$$
Let the output friendly sub-system has the form $x_y^+=Lu_1u_2x_y^1x_y^2x_r^1\xi_d^1$. Then in this system the effects of the disturbance can not removed, according to \cite{yang2013controller}. But, applying the \textit{Algorithm A} various sets are obtained as
$$S_1=\{2,3,4\},\ S_2=\{1\}$$
$$C_1=\{2,4\},\ C_2=\{2,4\},\ C_3=\{1,2\},\ C_4=\{1,2,3,4\}$$
This results in a total of $2^2\cdot 2^2\cdot 2^2\cdot 4^2=1024$ state feedback matrices that effectively remove the disturbance from the system, a possible one of which can be as
$$M_x=\delta_4[2\ 2\ 4\ 4\ 2\ 2\ 3\ 3]$$



\ifCLASSOPTIONcaptionsoff
  \newpage
\fi

%




\bibliographystyle{IEEEtran}
\bibliography{BCN_8Aug19}

%





\end{document}